\documentclass[%
 aip,
pop,
 amsmath,amssymb,
 reprint,%
]{revtex4-1}

\usepackage{graphicx}
\usepackage{dcolumn}
\usepackage{bm}
\usepackage[utf8]{inputenc}
\usepackage{todonotes}
\usepackage[colorlinks=true, allcolors=blue]{hyperref}
\usepackage{multirow}

\usepackage{scalerel,stackengine}
\stackMath
\newcommand\reallywidehat[1]{%
\savestack{\tmpbox}{\stretchto{%
  \scaleto{%
    \scalerel*[\widthof{\ensuremath{#1}}]{\kern-.6pt\bigwedge\kern-.6pt}%
    {\rule[-\textheight/2]{1ex}{\textheight}}
  }{\textheight}%
}{0.5ex}}%
\stackon[1pt]{#1}{\tmpbox}%
}

\newcommand{\V}[1]{\mathbf{#1}}
\newcommand{\Vshell}[2]{\mathbf{#1}^\mathrm{#2}}
\newcommand{\shell}[2]{{#1}^\mathrm{#2}}
\newcommand{\FT}[1]{\reallywidehat{#1}}

\newcommand{\T}[1]{\mathcal{T}_{\mathrm{#1}}}

\newcommand{\RE}[1]{\mathrm{Re} \left [ #1 \right ]}
\newcommand{\bra}[1]{\left ( #1 \right )}

\newcommand{\Ms}{\mathrm{M_s}}
\newcommand{\Ma}{\mathrm{M_a}}
\newcommand{\pd}{\partial}
\newcommand{\Emag}{E_\mathrm{B}}
\newcommand{\Ekin}{E_\mathrm{u}}

\newcommand{\Va}{\mathbf{v}_{\mathrm{A}}}
\newcommand{\imagi}{\mathrm{i}}
\newcommand{\Eulere}{\mathrm{e}}
\newcommand{\pb}{\beta_{\mathrm{p}}}
\newcommand{\cs}{c_{\mathrm{s}}}

\begin{document}

\title[Energy transfer in compressible MHD turbulence]{
Energy transfer in compressible magnetohydrodynamic turbulence
}

\author{Philipp Grete}
 \email{grete@pa.msu.edu.}
\affiliation{
Department of Physics and Astronomy,
Michigan State University, East Lansing, MI 48824, USA}
\author{Brian W. O'Shea}%
\affiliation{
Department of Physics and Astronomy,
Michigan State University, East Lansing, MI 48824, USA}
\affiliation{
Department of Computational Mathematics, Science and Engineering,
Michigan State University, East Lansing, MI 48824, USA}
\affiliation{
National Superconducting Cyclotron Laboratory, 
Michigan State University, East Lansing, MI 48824, USA}

\author{Kris Beckwith}
\affiliation{%
Tech-X Corporation, 5621 Arapahoe Ave. Boulder, CO, USA
}%

\author{Wolfram Schmidt}
\affiliation{
Hamburger Sternwarte, Universität Hamburg, Gojenbergsweg 112, D-21029 Hamburg, Germany
}

\author{Andrew Christlieb}
\affiliation{
Department of Computational Mathematics, Science and Engineering,
Michigan State University, East Lansing, MI 48824, USA}

\date{\today}

\begin{abstract}

Magnetic fields, compressibility and turbulence are important factors in many
terrestrial and astrophysical processes.
While energy dynamics, i.e. how energy is transferred within and between kinetic
and magnetic reservoirs, has been previously studied in the context of 
incompressible magnetohydrodynamic (MHD) turbulence, we extend
shell-to-shell energy transfer analysis to the compressible regime.
We derive four new transfer functions specifically capturing compressibility
effects in the kinetic and magnetic cascade, and capturing energy exchange via
magnetic pressure. 
To illustrate their viability, we perform and analyze four simulations of 
driven {isothermal} MHD turbulence in the sub- and
supersonic regime with two different codes.
On the one hand, our analysis reveals robust characteristics across regime and 
numerical method.
{For example, energy transfer between individual scales is local and forward
for both cascades with the magnetic cascade being stronger than the kinetic one.
Magnetic tension and magnetic pressure related transfers are less local
and weaker than the cascades.
We find no evidence for significant nonlocal transfer.
}
On the other hand, we show that certain functions, e.g., the compressive component
of the magnetic energy cascade, exhibit a more complex behavior {that varies
both with regime and numerical method}.
Having established a basis for the analysis in the compressible regime, the 
method can now be applied to study  a broader parameter space.
\end{abstract}

\maketitle

\section{Introduction}

Compressible, magnetized turbulence is thought to play an important
role in many astrophysical\cite{Brandenburg2013} and terrestrial processes.
These include the amplification of magnetic fields, e.g. in
dynamos\cite{Brandenburg2005,Tobias2013}, or 
particle acceleration in shocks as the origin of
cosmic rays\cite{Brunetti2015}.
However, our current understanding of turbulence even in the simplest 
description of a magnetized plasma, magnetohydrodynamics (MHD), 
is far from complete  and much less developed than
theories of incompressible hydrodynamic turbulence.

While the nonlinearities in the incompressible hydrodynamic regime lead
to the well known energy cascade and a well defined inertial range (in the absence 
of dissipative effects)\cite{Frisch1995}, the overall picture in 
MHD is more complex\cite{biskamp}.
Three ideal {quadratic} invariants exists {in incompressible MHD}: 
energy, cross-helicity and magnetic helicity.
Our present study concerns the energy dynamics only.
Non-trivial transfers
between and within kinetic and magnetic energy reservoirs are possible
even in the case of vanishing cross-helicity and magnetic helicity.

The inherent nonlinearities of the governing equations make an exact 
analytic treatment very challenging.
In {the incompressible regime}, these nonlinearities can be understood as triad 
interactions\cite{Kraichnan1967}, i.e.
energy at some scale is transferred to energy at a second scale via a 
mediating interaction at a third scale.
All these scales need to form a closed triangle in spectral space.
Understanding and quantifying these nonlinearities, e.g.
in experimental, observational and numerical data,
facilitates advances in turbulence research in the absence of a universal theory.
For example, results and conclusions from this kind of analysis can be used
in the development of subgrid-scale models for large eddy simulations.

Prior work\citep{Alexakis2005,Verma2004,Debliquy2005,Mininni2011}
examined the locality of energy transfers in incompressible MHD turbulence, with evidence presented for both local and (strong) non-local transfers and transfer between kinetic and magnetic reservoirs.
The origin of this discrepancy was eventually shown to be\cite{Aluie2010} uses of
different definitions of shells in spectral space over which the energy transfer
takes place. On the one hand, linear binning overestimates the influence of the 
largest scales; however, logarithmic binning allows for localized structures in
physical and spectral space and asymptotically favors local interactions.
This is closer to phenomenological descriptions and some groups revised
their earlier work to confirm the new interpretation of 
weakly local transfer\citep{Teaca2011}. This highlights the importance of using a well-defined formalism to interpret the non-linear dynamics of MHD energy cascades. One of the outcomes from this work is such a formalism for compressible MHD turbulence.

Such a formalism is necessary as the importance of understanding energy cascades in compressible MHD turbulence has recently become apparent in a range of applications. Total energy transfers, i.e. summing over all possible interaction with all scales,
has been analyzed in the context of small-scale dynamo action in solar
magneto-convection\citep{Graham2010}, of the magnetorotational instability (MRI)
in a shearing box\citep{FromangS.2007}, and of magnetized Kelvin-Helmholtz
instabilities\citep{Salvesen2014}. This work has been used to investigate important issues related to numerical convergence of angular momentum transport arising from the MRI\citep{FromangS.2007} and to understand dissipation of turbulent fluctuations\citep{Salvesen2014}. Beyond analysis of total energy transfers, cross-scale fluxes, i.e. the amount of energy being transferred from scales larger than a certain scale to smaller scales, has been analyzed in the
compressible regime\citep{Yang2016}, while a more detailed study\cite{Moll_2011} also analyzed shell-to-shell energy transfers in the compressible regime.
However, the transfer functions presented by these authors result in a single combined term representing the magnetic cascade and magnetic pressure interactions.

In this work we extend this approach, illustrating how shell-to-shell transfer 
functions can be calculated in the compressible regime separating magnetic cascade
dynamics from magnetic pressure dynamics. We apply the resulting formalism to an ensemble of driven compressible MHD turbulence simulations in the subsonic and (weakly) supersonic regimes. We use the results from these calculations to highlight similarities in the MHD turbulence that arises from {two} different numerical schemes, the role played by forcing in determining the turbulence cascade in the supersonic case and suggest important physics that need to captured by subgrid-scale models of MHD turbulence.

The rest of this paper is organized as follows.
In Section~\ref{sec:method}, we derive the energy transfer terms in compressible MHD
in the next section and introduce our numerical simulations.
In Section~\ref{sec:results} we present the results starting
from a high-level view on cross-scale and total energy transfer and
close with the individual shell-to-shell transfers.
Then, the results are discussed in section~\ref{sec:discussion} and put 
into context to other results.
Finally, we conclude the paper in section~\ref{sec:conclusions} where
potential future directions are highlighted.

\section{Methods}
\label{sec:method}
\subsection{Energy transfer in compressible MHD}
In general, we follow the presentation and notation used by 
Alexakis \textit{et al.}\cite{Alexakis2005} in
the incompressible MHD regime.
Fourier transforms are denoted by a $\FT{\Box}$ and are defined for an
arbitrary quantity $\phi$ as 
\begin{align}
\label{eq:FT}
\FT{\phi} \bra{\V{k}} &= \frac{1}{\bra{2\pi}^3} 
\int \phi \bra{\V{x}} \Eulere^{-\imagi \V{k} \cdot \V{x}} \mathrm{d}\V{x} \quad \text{and} \\
\label{eq:iFT}
\phi \bra{\V{x}} &= \int \FT{\V{\phi}} \bra{\V{k}} \Eulere^{\imagi \V{k}\cdot\V{x}} \mathrm{d}\V{k}\;.
\end{align}
Complex conjugates are indicated by a star ($\Box^*$).
Summation convention, i.e. summation over repeated indices applies to all formulas.
If not noted otherwise all real space quantities depend on $\V{x}$ and all spectral
space quantities depend on normalized, dimensionless wavenumber $\V{k}$.
\subsubsection{Energy equations}

We start with the compressible ideal MHD equations in conservative form
\begin{align}
\label{eq:rho}
\pd_t \rho + \pd_j \rho u_j &= 0 \;, \\
\label{eq:rhoU}
\pd_t \rho u_i + \pd_j \rho u_i u_j - \pd_j B_i B_j + \pd_i p +\pd_i B_j B_j /2 &= f_i \;, \\
\label{eq:B}
\pd_t B_i -  \pd_j u_i B_j + \pd_j u_j B_i  &= 0 \;.
\end{align}
The density is denoted by $\rho$, the velocities by $\V{u}$, and the thermal pressure
by $p$.
The magnetic field $\V{B}$ incorporates a factor $1/\sqrt{4\pi}$.
$\V{f}$ on the right hand side of the momentum equation represents an external force.
For our simulations of mechanically forced turbulence it is given by an acceleration field 
(see section \ref{sec:numericalData}) $a_i$ with $f_i = \rho a_i$.
{The system is closed with an isothermal equation of state.} 

In order to speak of scale interactions in energy transfer, we need a definition of
the kinetic and magnetic energy in wavenumber space.
The definitions of the magnetic energy densities ($\Emag$) are straightforward by virtue
of Parseval's theorem for the total magnetic energy
\begin{align}
\int \underbrace{\frac{1}{2} B_i B_i}_{\equiv \Emag(\V{x})} \mathrm{d}\V{x} = 
\frac{1}{\bra{2\pi}^3} \int \underbrace{\frac{1}{2}\FT{B}_i \FT{B}_i^*}_{\equiv \Emag(\V{k})} 
\mathrm{d}\V{k}
\;.
\end{align}
The dynamic equations can simply be derived by multiplying \eqref{eq:B} with 
$B_i$
\begin{align}
\label{eq:MagEnReal}
\pd_t \Emag \bra{\V{x}} =  B_i \pd_j u_i B_j - B_i \pd_j u_j B_i
\;,
\end{align}
or the Fourier transform of \eqref{eq:B} with $\FT{B}_i^*$, respectively
\begin{align}
\label{eq:MagEnFT}
\pd_t \Emag \bra{\V{k}} = \RE{ \FT{B}_i \FT{\pd_j u_i B_j}^* - \FT{B_i} \FT{\pd_j u_j B_i}^*}
\;.
\end{align}

The spectral kinetic energy densities ($\Ekin$) in compressible (M)HD are not unique.
Two different versions are commonly used.
The first option is based on mixed complex conjugates, i.e. 
$\Ekin(\V{k}) = \RE{\FT{u}_i \FT{\rho u}_i^*}/2$ and used for example in
\cite{Graham2010,Salvesen2014}.
However, this definition does not guarantee positive definiteness of the energy in
wavenumber space.
For this reason, we follow Kida \& Orszag\cite{Kida1990} and introduce a new quantity
\begin{align}
\V{w} \equiv \sqrt{\rho} \V{u}.
\end{align}
It can be see as an analogous expression to the magnetic field 
$\V{B}=\sqrt{\rho}\Va$ in terms of the Alf\'en velocity
$\Va$. 
The kinetic expression allows for a positive definite definition of the 
kinetic energy density 
\begin{align}
\int \underbrace{\frac{1}{2} w_i w_i}_{\equiv \Ekin(\V{x})} \mathrm{d}\V{x} = 
\frac{1}{\bra{2\pi}^3} \int \underbrace{\frac{1}{2} \FT{w}_i \FT{w}_i^*}_{\equiv \Ekin(\V{k})} 
\mathrm{d}\V{k}
\;.
\end{align}
Here, the derivation of the dynamic equations requires an additional step.
Given that
\begin{align}
\pd_t w_i = u_i \pd_t \sqrt{\rho} + \sqrt{\rho}\pd_t u_i
\end{align}
we  can rewrite \eqref{eq:rho} to
\begin{align}
\pd_t \sqrt{\rho} = - \frac{1}{2\sqrt{\rho}} \pd_j \rho u_j = 
- \frac{1}{2} \sqrt{\rho} \pd_j u_j - u_j \pd_j \sqrt{\rho}
\end{align}
and \eqref{eq:rhoU} to
\begin{align}
\pd_t u_i = - u_j \pd_j u_i  + \frac{1}{\rho} \pd_j B_i B_j - \frac{1}{\rho} \pd_i p - \frac{1}{2\rho} \pd_i B_j B_j + a_i\;,
\end{align}
in order to obtain the dynamical equation for $\V{w}$
\begin{align}
\begin{split}
\pd_t w_i = &- u_j \pd_j w_i  - \frac{1}{2} w_i \pd_j u_j
+ \frac{1}{\sqrt{\rho}} \pd_j B_i B_j \\
&- \frac{1}{\sqrt{\rho}} \pd_i p - \frac{1}{2 \sqrt{\rho}} \pd_i B_j B_j + \sqrt{\rho} a_i\;.
\end{split}
\end{align}
\begin{widetext}
\noindent With this equation it is now possible to write write down the dynamical equations for the kinetic energy density analogous to the magnetic case.
The dynamic equation in real space is
\begin{align}
\label{eq:KinEnReal}
\pd_t \Ekin (\V{x}) = &- w_i u_j \pd_j w_i  - \frac{1}{2} w_i w_i \pd_j u_j
+ \frac{w_i}{\sqrt{\rho}} \pd_j B_i B_j
- \frac{w_i}{\sqrt{\rho}} \pd_i p - \frac{w_i}{2 \sqrt{\rho}} \pd_i B_j B_j + w_i \sqrt{\rho} a_i,
\end{align}
and in wavenumber space
\begin{align}
\label{eq:KinEnFT}
\pd_t \Ekin (\V{k}) =  \RE{ - \FT{w_i} \FT{u_j \pd_j w_i}^*  - \frac{1}{2} \FT{w_i} \FT{w_i \pd_j u_j}^* 
+ \FT{w_i} \FT{\frac{1}{\sqrt{\rho}} \pd_j B_i B_j}^* - \FT{w_i} \FT{\frac{1}{\sqrt{\rho}} \pd_i p}^* 
- \FT{w_i} \FT{\frac{1}{2 \sqrt{\rho}} \pd_i B_j B_j}^* +
\FT{w_i} \FT{\sqrt{\rho} a_i}^*
}\;.
\end{align}
It should be noted that the expression in real space is equivalent to the standard definition based on 
$\pd_t \rho u_i u_i$ (c.f. Eqn. 28 - 35 of Simon \textit{et al.}\cite{Simon2009})
\end{widetext}

\subsubsection{Expressing energy equations in terms of interactions}
Starting from the energy equations \eqref{eq:MagEnFT} and \eqref{eq:KinEnFT}
for a single wavenumber $\V{k}$ we now illustrate how to break these down to 
individual interacting scales.
Given that we analyze isotropic turbulence, individual wavenumbers are of less interest than collective behavior within shells.
For this reason we define the shell-filtered quantities in real space as
\begin{align}
\label{eq:DefShell}
\shell{\phi}{K}\bra{\V{x}} = 
\int_{\mathrm{K}} \FT{\V{\phi}} \bra{\V{k}} \Eulere^{\imagi \V{k}\cdot\V{x}} 
\mathrm{d}\V{k}
\;.
\end{align}
where the integration over $\mathrm{K}$ stands for the integration over shell
$\mathrm{K}$.
It should not be confused with a specific wavenumber.
Different definitions of the actual shells $\mathrm{K}$ exist in the literature.
Given that the choice of the shells itself is independent of the
derivation of the transfer functions, 
we defer a more detailed discussion of different definitions to the next 
subsection~(\ref{sec:binning}).
Note that we use the terms bin and shell interchangeably in the following sections.

Naturally, the summation over all shells recovers the original field
\begin{align}
\label{eq:fullShell}
\phi\bra{\V{x}} = \sum_\mathrm{K} \shell{\phi}{K}\bra{\V{x}} \;.
\end{align}
With these definitions we can now illustrate the derivation of shell-to-shell
transfer terms for one sample term -- the first term in \eqref{eq:KinEnFT}:
\begin{align}
\RE{ - \FT{w_i}\bra{\V{k}} \FT{u_j \pd_j w_i}^*}\bra{\V{k}}  \;.
\end{align}
First, we replace the last $w_i$ in the equation with its shell decomposed definition
according to \eqref{eq:fullShell}.
Then, we explicity write down the Fourier transform \eqref{eq:FT}
of the complex conjugate term and get
\begin{align}
\RE{ - \FT{w_i}\bra{\V{k}} 
\int \Eulere^{\imagi \V{k}\cdot\V{x}} u_j \pd_j \sum_\mathrm{Q} \shell{w_i}{Q} \mathrm{d}\V{x}} \;.
\end{align}
Given that the integration itself is independent of $k$ we can pull 
$\FT{w_i}\bra{\V{k}}$ into the integration.
Similarly, the summation can be rearranged as individual shells that are
orthogonal to each other, resulting in
\begin{align}
\int \sum_\mathrm{Q}  \RE{ - \FT{w_i}\bra{\V{k}} 
\Eulere^{\imagi \V{k}\cdot\V{x}} u_j \pd_j \shell{w_i}{Q}} \mathrm{d}\V{x} \;.
 \end{align}

Now, the final step is to remember that we are interested in the evolution within shells
rather than individual modes $k$.
Thus,
{defining the kinetic energy in a shell $\mathrm{K}$
as 
$\Ekin^\mathrm{K} = \int_\mathrm{K} \Ekin (\V{k}) \mathrm{d}\V{k}$
}
and using \eqref{eq:DefShell} 
we can write
the change of kinetic energy in that shell
as
\begin{align}
\label{eq:totalTransExample}
\pd_t \Ekin^\mathrm{K} = \int \sum_\mathrm{Q} - \shell{w_i}{K} u_j \pd_j \shell{w_i}{Q} \mathrm{d}\V{x} + \ldots
\end{align}
This procedure similarly applies to all other terms in the energy equations
and the resulting terms for the compressible ideal MHD equations are given in the 
following subsection.

\subsubsection{Transfer functions}
\begin{figure}[htbp]
\centering
\includegraphics[width=0.8\columnwidth]{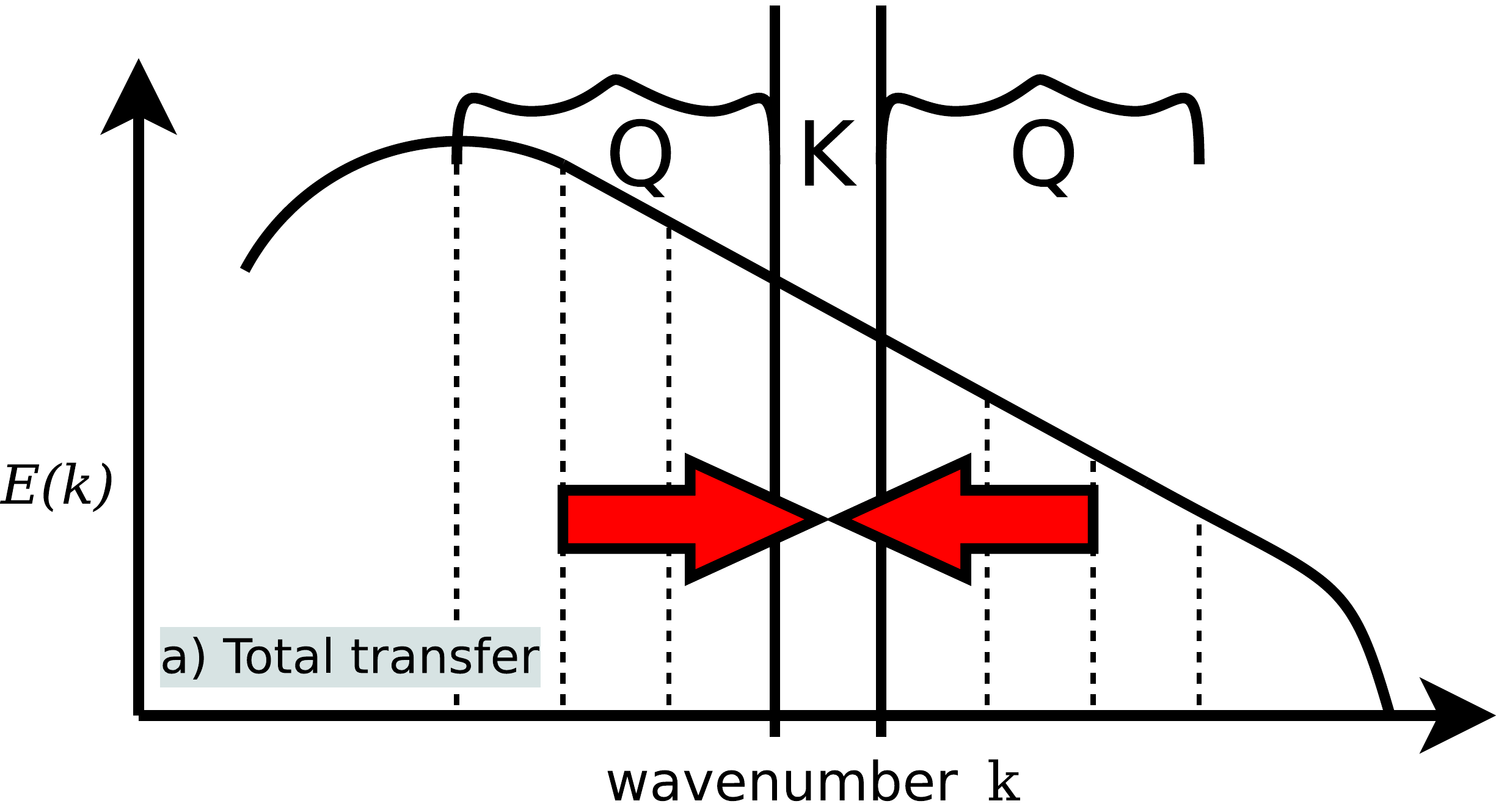}
\includegraphics[width=0.8\columnwidth]{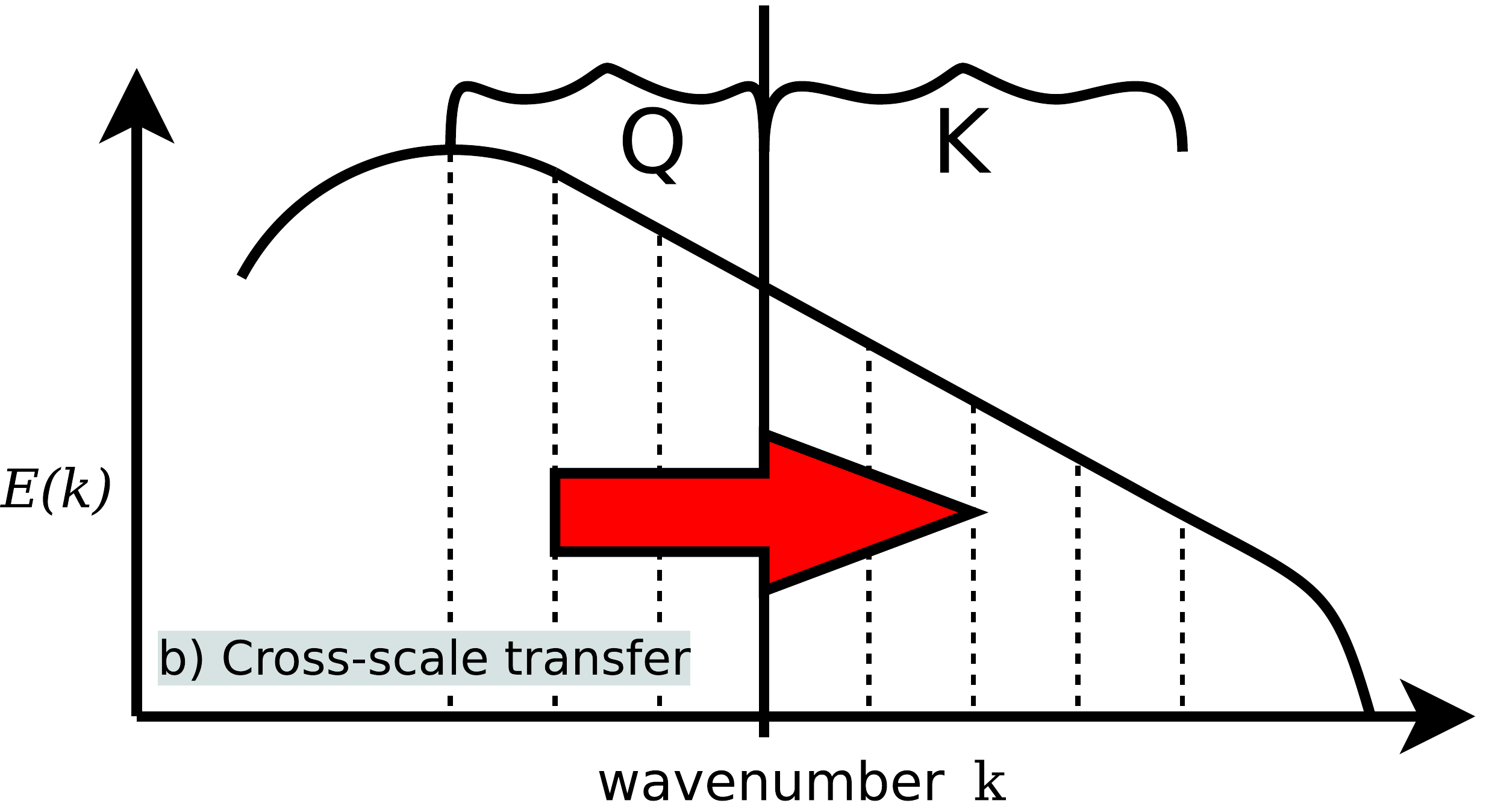}
\includegraphics[width=0.8\columnwidth]{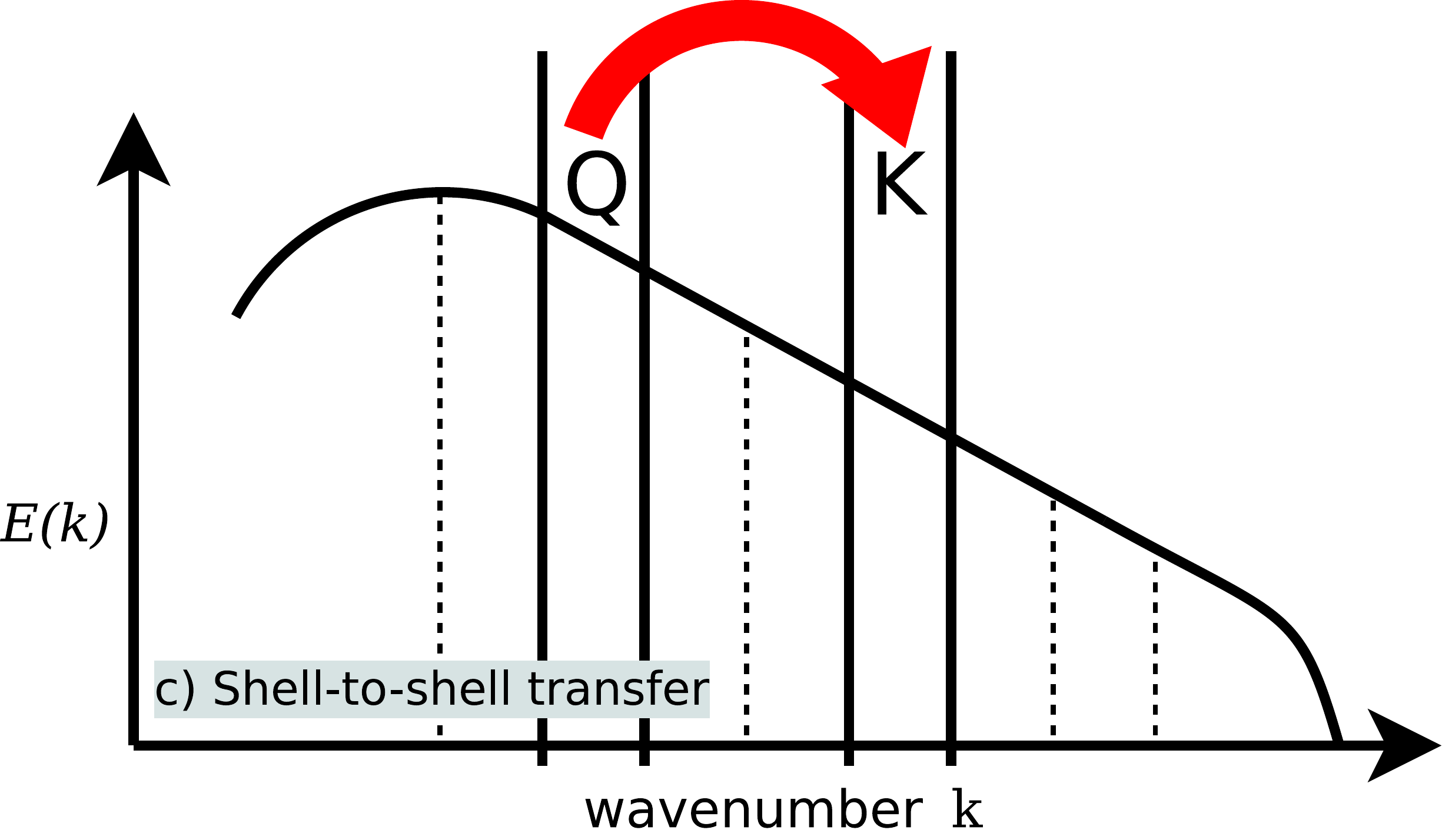}
\caption{Illustrations of the energy transfers that are analyzed 
in this paper: a) the total transfer, i.e. energy from all
shells going to a particular shell (see \ref{sec:ResTotal}),
b) cross-scale transfer, i.e. energy from all scales larger than a particular
scale going to all scales smaller than that scale (see 
\ref{sec:DefCross} and \ref{sec:ResCrossScale}), and
c) shell-to-shell transfer, i.e. energy going from a particular shell $Q$
to another shell $K$ (see \ref{sec:ResShell}).
Naturally, all shells can be either part of the kinetic or magnetic energy
reservoir, which is not further illustrated in the sketch.}
\label{fig:sketches}
\end{figure}
The total transfer in (or out) of a shell $\mathrm{K}$ is given 
by~\eqref{eq:totalTransExample} and illustrated in Fig.\ref{fig:sketches}a.
It can be further separated by looking at an individual shell $\mathrm{Q}$ 
rather than the sum over all shells
as shown in Fig.\ref{fig:sketches}c.
In the following, we denote individual transfers by
\begin{align}
\T{XY} (Q,K) \quad \text{with} \quad \mathrm{X,Y} \in \{\mathrm{U,B} \}
\end{align}
expressing energy transfer (for $\T{} > 0$) 
from shell $\mathrm{Q}$ of energy reservoir 
$\mathrm{X}$ to shell $\mathrm{K}$ of energy reservoir 
$\mathrm{Y}$.
The kinetic energy reservoir identified by $\mathrm{U}$ is always linked 
to a shell filtered $\V{w}$ quantity, whereas the magnetic energy reservoir
identified  by $\mathrm{B}$ is always linked to a shell filtered $\V{B}$ quantity.
If a third lower index is present it refers to the mediating force and is not
linked to a specific reservoir.

In general, we define all fundamental transfers so that they satisfy antisymmetry
\begin{align}
\T{XY} (Q,K) = - \T{YX} (K,Q)\;.
\end{align}
In other words, energy transferred from shell $\mathrm{Q}$ of reservoir 
$\mathrm{X}$ to shell $\mathrm{K}$ of reservoir 
$\mathrm{Y}$ is, by definition, equal to the amount of energy received
by shell  $\mathrm{K}$ of reservoir  $\mathrm{Y}$ from shell
$\mathrm{Q}$ of reservoir $\mathrm{X}$.

It should be noted that the formalism used here allows one to draw conclusions
about the transfer of energy between scales, but not whether this transfer
is local or non-local with respect to the mediating mode.
From the point of view of triad interactions we generally do not restrict 
the third, mediating quantity in the transfer functions.
For this reason, conclusions on the nature of the interaction -- for example, whether
it is local or non-local -- cannot be easily drawn from analyzing these terms,
especially if linearly binned shells are used~\citep{Alexakis2005,Aluie2009}.

\paragraph{Kinetic cascade terms}
The first two terms in the kinetic energy equation \eqref{eq:KinEnFT} correspond
to transfer attributed to the kinetic cascade as they mediate energy transfer
within the kinetic energy reservoir. 
They are given by
\begin{align}
\T{UU}(Q,K) = - \int \underbrace{\Vshell{w}{K} \cdot \bra{\V{u} \cdot \nabla} \Vshell{w}{Q}}_{\equiv\T{UUa}}  +
\underbrace{\frac{1}{2} \Vshell{w}{K} \cdot \Vshell{w}{Q} \nabla \cdot \V{u}}_{\equiv \T{UUc}}
\mathrm{d}\V{x} \;.
\end{align}
The total term can be further split into an advective component, $\T{UUa}$, and
a compressive component, $\T{UUc}$.
The former is equivalently present in incompressible MHD, whereas the
latter explicitly captures compressive dynamics.
It should be noted, that only the total term $\T{UU}$ satisfies antisymmetry
and not the individual terms.

\paragraph{Magnetic tension related terms}
The third term in the kinetic energy equation \eqref{eq:KinEnFT} regulates energy
transfer from magnetic energy to kinetic energy by magnetic tension.  This term in the 
equation describes energy transfer in the direction of the Alf\'en wave propagation through
tension, and it is given by
\begin{align}
\T{BUT}(Q,K) = \int \Vshell{w}{K} \cdot  \bra{ \Va \cdot \nabla} \Vshell{B}{Q}  \mathrm{d}\V{x} \;.
\end{align}
Here, we applied $\nabla\cdot\Vshell{B}{Q}=0$.

It is antisymmetric to the corresponding first term in the magnetic energy 
equation~\eqref{eq:MagEnFT}, 
\begin{align}
\T{UBT}(Q,K) =  \int  \Vshell{B}{K} \cdot \nabla \cdot \bra{\Va \otimes \Vshell{w}{Q}} \mathrm{d}\V{x},
\end{align}
where $\otimes$ denotes a tensor product.
By nature of the symmetry, $\T{UBT}$ transfers energy from the kinetic energy reservoir to
the magnetic one.
Here, we replaced the magnetic field and the velocity field by the Alf\'en velocity and
the density weighted velocity, respectively.
In addition to satisfying symmetry, this replacement allows for a consistent treatment
of the kinetic energy reservoir.
In the compressible regime, terms related to normal shell filtered 
velocities, e.g. $\Vshell{u}{Q}$, correspond to the specific 
kinetic energy.
However, we are interested in the dynamics of the {energy densities} and
therefore introduce/replace the appropriate $\Vshell{w}{Q}$ where necessary ---
also in the following equations.

\paragraph{Magnetic pressure and cascade related terms}
\label{sec:DefBBandBUP}

The last term in the kinetic energy equation \eqref{eq:KinEnFT}
corresponds to changes due to magnetic pressure. 
We treat this term in a fashion analogous to that proposed by 
Fromang \& Papaloizou\cite{FromangS.2007} and Simon \textit{et al.}\cite{Simon2009}, 
which allows the magnetic pressure to be decoupled from the magnetic cascade dynamics 
(contrary to the suggestion of Moll \textit{et al.}\cite{Moll_2011}). 
We proceed by applying
shell decomposition directly to the last term in equation \eqref{eq:KinEnFT}:
\begin{align}
\label{eq:TBUPwBB}
\T{BUP}(Q,K) = - \int \frac{\Vshell{w}{K}}{2 \sqrt{\rho}} \cdot \nabla \bra{ \V{B} \cdot \Vshell{B}{Q}}  \mathrm{d}\V{x} \;.
\end{align}
In other words, if one factor $B_j$ is associated with a shell Q, resulting in $\shell{B}{Q}_j$, and the other with all mediating modes, $\sum_\mathrm{P} \shell{B}{P}_j = B_j$, the product rule for quadratic expressions does not apply to $\shell{B}{P}_j\shell{B}{Q}_j$ such that the magnetic to kinetic transfer is given by the above expression. In order to derive the corresponding (antisymmetric) transfer term in the magnetic energy equation, we first expand the original term to
\begin{multline}
\RE{- \FT{B_i} \FT{\pd_j u_j B_i}^*}  =  
\RE{- \FT{B_i} \FT{u_j \pd_j B_i}^* - \FT{B_i} \FT{B_i \pd_j u_j}^* }\\
\label{eq:MagTermMagPres}
 = \RE{- \FT{B_i} \FT{u_j \pd_j B_i}^* - 
 \frac{1}{2} \FT{B_i} \FT{B_i \pd_j u_j}^* - 
 \FT{ B_i} \FT{B_i \pd_j \frac{w_j}{2 \sqrt{\rho}}}^*}  \;.
\end{multline}
The last term in \eqref{eq:MagTermMagPres} can be associated with
transfer from kinetic energy to magnetic energy via magnetic pressure
\begin{align}
\label{eq:TUBPwBB}
\T{UBP}(Q,K) = - \int \Vshell{B}{K} \cdot \V{B}  \nabla \cdot \bra{\frac{\Vshell{w}{Q}}{2 \sqrt{\rho}}}  \mathrm{d}\V{x} \;.
\end{align}
Again, this term satisfies antisymmetry with its counterpart \eqref{eq:TBUPwBB} within this formulation.

The first two terms in  \eqref{eq:MagTermMagPres}  can now be associated with magnetic 
to magnetic transfer, i.e. a magnetic cascade with
\begin{align}
\T{BB}(Q,K) = - \int \underbrace{\Vshell{B}{K} \cdot \bra{\V{u} \cdot \nabla} \Vshell{B}{Q}}_{\equiv\T{BBa}}  +
\underbrace{\frac{1}{2} \Vshell{B}{K} \cdot \Vshell{B}{Q} \nabla \cdot \V{u}}_{\equiv \T{BBc}}
\mathrm{d}\V{x} \;.
\end{align}
This term satisfies antisymmetry with itself.
Moreover it has the identical shape as $\T{UU}$ in the kinetic energy equation
and can similarly be split into an advection-related component, $\T{BBa}$,
and a compression-related component, $\T{BBc}$.
Again, the advection term is already known from the incompressible MHD regime,
whereas the compressive term is new.

{It should be noted, that the last two terms in \eqref{eq:MagTermMagPres}
are mathematically identical and that the separation into magnetic pressure
and compression stems from chosen shell decomposition of the different
variables.
This hints at the dual nature of that term and represents our view on
separating magnetic cascade dynamics from the energy transfer between
kinetic and magnetic budgets as discussed later.
} 

\paragraph{Pressure and external force terms}
The two remaining terms in the kinetic energy equation \eqref{eq:KinEnFT}
are not associated with energy transfer between or within
kinetic and magnetic energy reservoirs.

First, the pressure gradient term  is given by
\begin{align}
	\T{PU}(Q,K) = - \int \frac{1}{\sqrt{\rho}} \Vshell{w}{K}\cdot \nabla \shell{p}{Q}
	\mathrm{d}\V{x} \;.
\end{align}
It allows for an exchange of energy between the kinetic reservoir and the internal
energy reservoir, or, in case of isothermal turbulence, density fluctuations.
{Given that the present manuscript is primarily concerned with the dynamics
within and betweeen kinetic and magnetic energy budgets and that we analyze 
sub- and mildly supersonic, isothermal turbulence, this term and internal
energy dynamics are generally not discussed in great detail.}

Second, energy is injected by a mechanical force.
The exact shell-to-shell transfer of that external force is given by
\begin{align}
	\label{eq:FU}
	\T{FU}(Q,K) = - \int \sqrt{\rho} \Vshell{w}{K}\cdot \Vshell{a}{Q}
	\mathrm{d}\V{x} \;.
\end{align}
It is easily seen that if the force is specified via an acceleration field 
the density field is acting as a mediator.
Some implications of this are discussed in a later section.

\paragraph{Summary}
Putting all terms together the interplay of kinetic and magnetic
energies in a shell $\mathrm{K}$ is given by
\begin{align}
	\begin{split}
\pd_t \Ekin^\mathrm{K} = \int \sum_\mathrm{Q} \Big( & 
\T{UUa} + \T{BUT} + \T{PU} + \T{FU}  + \\
&
\textcolor{blue}{\T{UUc}} + \textcolor{blue}{\T{BUP}} \Big)
\; \mathrm{d} \V{x} + \mathcal{D}
\end{split}\\
\pd_t \Emag^{\mathrm{K}} = \int \sum_\mathrm{Q} \Big( &
\T{BBa} + \textcolor{blue}{\T{BBc}} + \T{UBT} + \textcolor{blue}{\T{UBP}} \Big)
\; \mathrm{d} \V{x} 
+ \mathcal{D} 
\end{align}
where the new terms that enter the formalism in the compressible 
regime are highlighted in \textcolor{blue}{blue}.
For completeness, we also indicate the numerical dissipation
present in the type of simulations we are analyzing by 
$\mathcal{D}$.

\subsection{Definition of shells}
\label{sec:binning}
\begin{figure*}[!htbp]
\centering
\includegraphics[width=\textwidth]{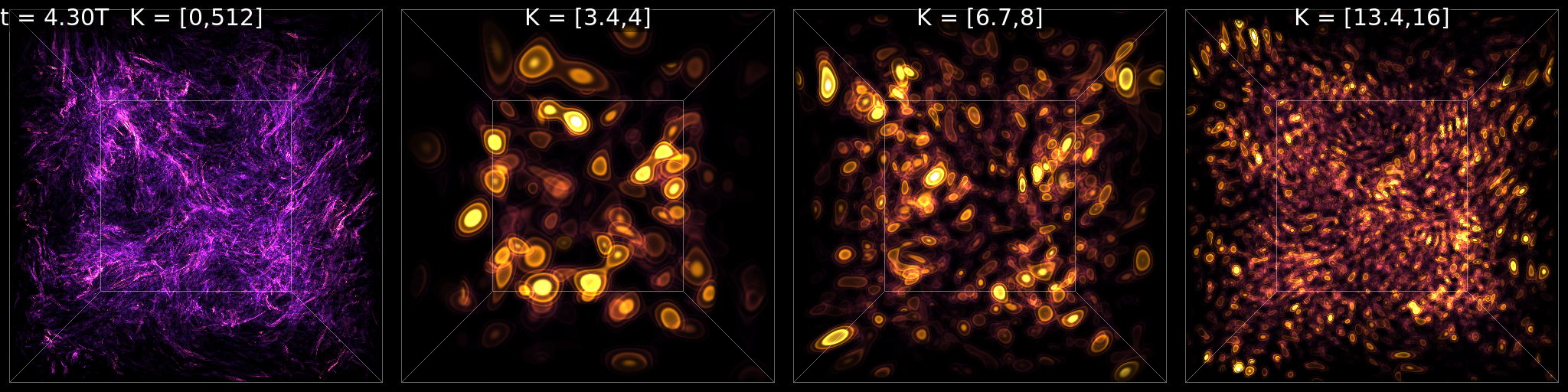}
\caption{Illustration of eddies of different sizes
in real space for the logarithmic binning employed in this paper.
Eddies are illustrated based on the \textit{Q-criterion} \cite{Hunt1988}, where
$Q =  \bra{\left\lVert \Omega \right\rVert^2 - \left\lVert\mathcal{S}\right\rVert^2}/2$ 
indicates regions dominated
by straining motion for $Q < 0$ and regions dominated by rotational motion for
$Q > 0$. To illustrate eddies we exclude regions with $Q$ smaller than $0.2$ times
its standard deviation.
Logarithmic binning allows for localized structures in real space.
\label{fig:BinningIllus}
}
\end{figure*}
The definition of the shells in spectral space is an important aspect 
of this kind of analysis.
Different
definitions probe different features, particularly if the locality of energy transfer is of concern.
For example, linear binning with 
$K \equiv k \in (\mathrm{K} - 0.5, \mathrm{K} + 0.5]$ 
is used\citep{Alexakis2005,Moll_2011} and corresponds to space-filling,
monochromatic wave-like structures.
Another example is octave binning with $K \equiv k \in (K/2, K]$ or 
more generally logarithmic binning\citep{CARATI2006,Aluie2009},
which allows for structures, such as eddies, that are simultaneously localized
in real and spectral space as illustrated in Fig.~\ref{fig:BinningIllus}.
Such an approach allows for more physically intuitive interpretations of the 
turbulent cascade to emerge as it is closer to the phenomenological cascade 
picture and naturally related to a power-law spectrum.
We adopt it here for these reasons.

\begin{figure}[!htbp]
\centering
\includegraphics{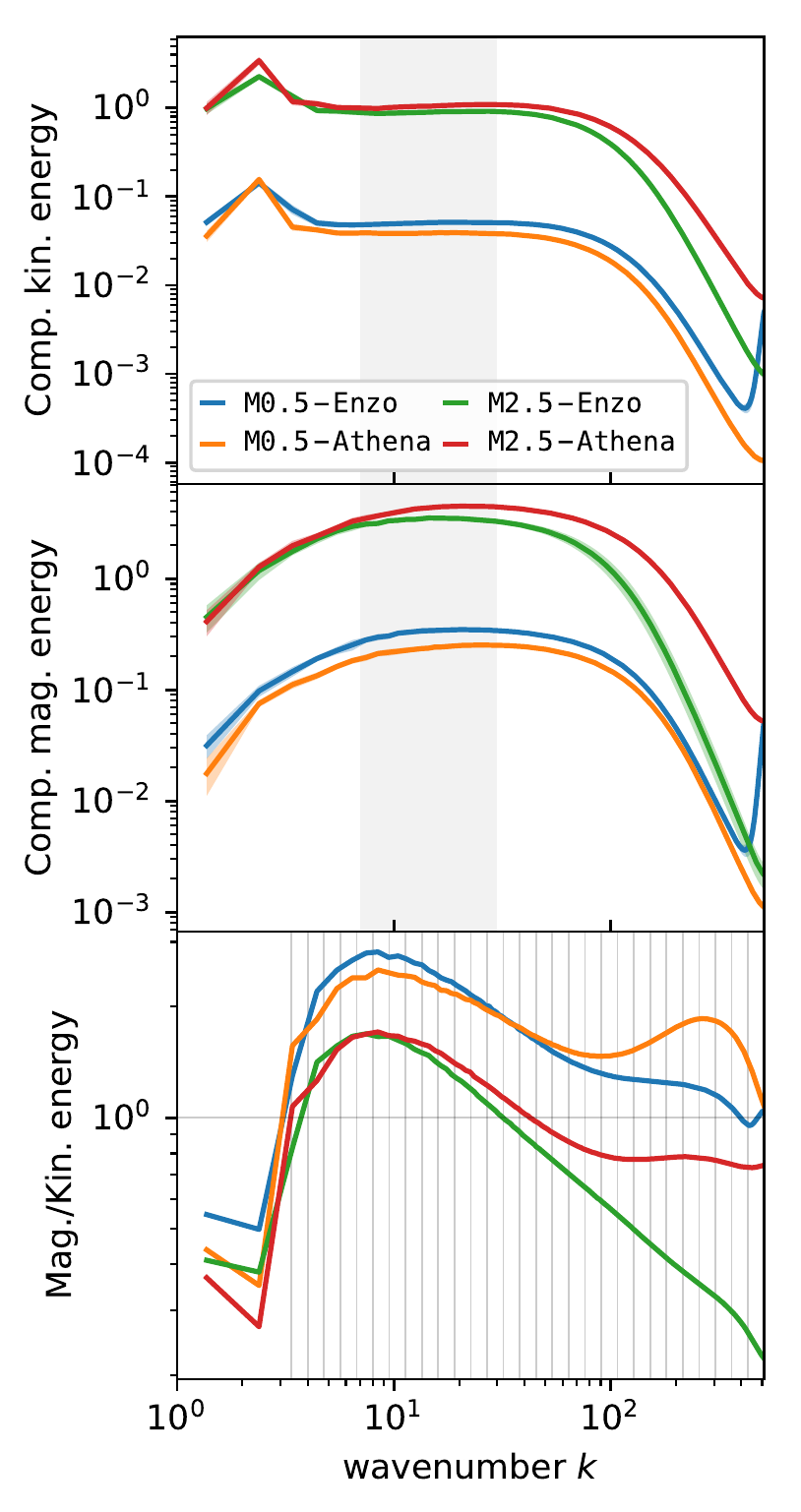}
\caption{
Compensated mean kinetic (top) and magnetic (center) energy spectra over all 
snapshots of a simulation.
The vertical transparent regions around each line illustrate the standard deviation.
For most of the lines the variation is very small and the transparent region lies
within the linewidth.
Grey areas highlight the range of scales ($7 < k < 30$) where approximately
power-law scaling is observed.
The kinetic energy is compensated by $k^{4/3}$ and the magnetic energy by $k^{1.7}$.
The scale-by-scale ratio (uncompensated) of magnetic to kinetic energy is shown 
in the bottom panel.
Vertical lines in the bottom panel illustrate the binning used throughout the paper.
\label{fig:Specs}
}
\end{figure}

In particular, our shell boundaries are given by $1$ and 
$2^{n/4 + 2}$ for $n \in \{ -1,0,1,\ldots,28\}$.
They are illustrated by the vertical lines in the bottom panel 
of Fig.~\ref{fig:Specs}.

In order to maintain a close relationship between wavenumber $k$ (in
lowercase letters) and shells $K$ (in uppercase letters) we identify shells
by the wavenumber they contain.
For example, $K=10$ refers to the shell containing $k=10$, i.e.
$k \in (9.5, 11.31]$.

\subsection{Cross-scale energy fluxes}
\label{sec:DefCross}
For the description of cross-scale energy fluxes, i.e. fluxes
from scales larger than a certain scale $k$ to the scales smaller
than $k$ as illustrated in Fig.~\ref{fig:sketches}b, we closely follow the exposition of \cite{Debliquy2005}
and extend it to the compressible case.

Based on the shell-to-shell energy transfer functions, the forward 
(large to small scales) cross-scale fluxes are generally given by
\begin{align}
\Pi^{\mathrm{X}^<}_{\mathrm{Y}^>} (k) = \sum_{Q \leq k} \sum_{K > k} \T{XY}(Q,K)
\end{align}
where $\mathrm{X,Y} \in \{\mathrm{U,B}\}$ again.
$\mathrm{X}^<$ refers to energy in reservoir $\mathrm{X}$ at wavenumbers
smaller than $k$, whereas 
$\mathrm{Y}^>$ refers to energy in reservoir $\mathrm{Y}$ at wavenumbers
larger than $k$.
\begin{figure}[!htbp]
\centering
\includegraphics[width=\columnwidth]{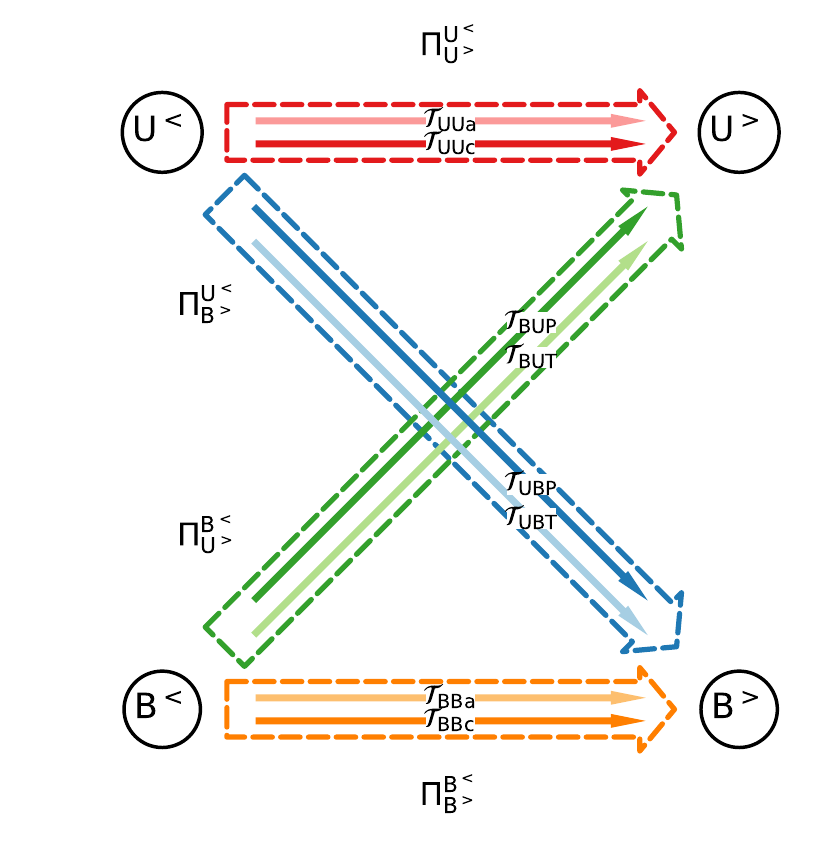}
\caption{Sketch of the cross-scale energy fluxes between
kinetic and magnetic energy reservoirs in
ideal compressible MHD.
The left spheres correspond to scales larger than a certain scale $k$,
whereas the right spheres correspond to smaller scales.
Each arrow illustrates transfer by a particular term.
Energy fluxes between different reservoirs are mediated by magnetic
pressure and tension forces and therefore have two arrows.
The $\Pi^{\mathrm{B}^<}_{\mathrm{B}^>}$ flux (orange arrow) is not present
when using variant \textrm{I} of the magnetic pressure formulation.
\label{fig:CrossScaleSketch}
}
\end{figure}
As in incompressible MHD
there are four different cross-scale fluxes in the compressible case between 
the kinetic and magnetic energy reservoirs as illustrated in
Fig.~\ref{fig:CrossScaleSketch}.
Contrary to incompressible MHD, however, they are not based on four different
transfer terms.
Here, we have one additional term for each flux.
For the intra-reservoir fluxes these are $\T{UUc}$ and $\T{BBc}$ specifically
capturing compressibility effects via $\nabla \cdot \V{u}$. 
For the inter-reservoir fluxes the additional terms related to magnetic 
pressure $\T{BUP}$ and $\T{UBP}$ enter the cross-scale fluxes.
They are given as 
\begin{align}
\Pi^{\mathrm{U}^<}_{\mathrm{U}^>} (k) &= 
\sum_{Q \leq k} \sum_{K > k} \T{UUa}(Q,K) + \T{UUc}(Q,K)\;, \\
\Pi^{\mathrm{B}^<}_{\mathrm{U}^>} (k) &= 
\sum_{Q \leq k} \sum_{K > k} \T{BUT}(Q,K) + \T{BUP}(Q,K) \;, \\
\Pi^{\mathrm{U}^<}_{\mathrm{B}^>} (k) &= 
\sum_{Q \leq k} \sum_{K > k} \T{UBT}(Q,K) + \T{UBP}(Q,K) \;, \\
\label{eq:CrossScaleFluxBB}
\Pi^{\mathrm{B}^<}_{\mathrm{B}^>} (k) &= 
\sum_{Q \leq k} \sum_{K > k} \T{BBa}(Q,K) + \T{BBc}(Q,K) \;.
\end{align}

\subsection{Numerical data}
\label{sec:numericalData}
\begin{table*}[!htbp]
        \begin{tabular}{lcccccc}
                \toprule
		Run  & Integrator & Riemann solver & $\nabla \cdot \V{B}$ & Isothermal EOS 
		& Forcing type & Forcing modes

                \\
\colrule

$\mathtt{M0.5-Enzo}$ & MUSCL-Hancock & HLLD & Cleaning & approx. & dynamic & mixed \\
$\mathtt{M0.5-Athena}$ & VL & HLLD & CT & exact & random & solenoidal \\
$\mathtt{M2.5-Enzo}$ & MUSCL-Hancock & HLL & Cleaning & approx. & dynamic & mixed \\
$\mathtt{M2.5-Athena}$ & VL & HLLD & CT & exact & random & solenoidal \\
\botrule
        \end{tabular}
        \caption{Overview of the numerical setup for each simulation.
		More details (including references) are given in the text.
        \label{tab:SimOverviewNum}} 
\end{table*}

We analyze four datasets of forced, homogeneous, isotropic MHD turbulence.
For the purpose of this paper, the datasets can be categorized
twofold: by regime and by numerical simulation code (see overview in 
Table~\ref{tab:SimOverviewNum}).
The regime is either approximately incompressible with a subsonic root-mean-square (r.m.s.)
Mach number of $\Ms \approx 0.5$, or weakly compressible with a supersonic r.m.s.~Mach
number of $\Ms \approx 2.5$.
In both regimes, we use the numerical codes 
\textsc{Enzo}\citep{Enzo2013}
and 
\textsc{Athena}\citep{Athena2008}
in
order to obtain physically similar simulations that differ by the numerical
method employed.
In all simulations we solve the ideal, compressible MHD equations on a static, periodic
grid with $1024^3$ points and side length of $1$ in code units.

The \textsc{Enzo} simulations were used before
in \textit{a priori} testing of subgrid-scale models\citep{Grete2016a}.
They use the MUSCL-Hancock framework with second-order Runge-Kutta 
time integration and the HLLD (in the subsonic regime) or HLL (in the supersonic
regime) Riemann solver\citep{Wang2009}.
The gas is kept approximately isothermal by using an adiabatic equation of state
with $\gamma = 1.001$.
The divergence constraint $\nabla \cdot \V{B} = 0$ is maintained by hyperbolic 
divergence cleaning\citep{Dedner2002}.

In the \textsc{Athena} simulations we employ the second-order van Leer 
integrator\citep{Stone2009} with HLLD Riemann solver in both regimes
and first-order flux correction.
Here, we use an exact isothermal equation of state.
Moreover, $\nabla \cdot \V{B} = 0$ is maintained to machine precision
by using constrained transport.

All simulations start with uniform initial conditions, i.e.,
$\rho_0 = 1$, $\V{u}_0 = \V{0}$, and $\V{B}_0 = \bra{0,0,\sqrt{2\rho_0 \cs^2 / \pb}}$,
with the speed of sound $\cs = 1$ in code units.
The initial plasma beta is $\pb = 72$ in the subsonic  regime and $\pb = 5$ 
in the supersonic regime, so that turbulence is super-Alf\'enic in the 
stationary regime.

Stationary turbulence is reached by constantly forcing the simulation box.
We employ an acceleration field with a parabolic shape in spectral space.
It peaks at low wavenumbers $k_0 = 2$ and is constrained to modes with
$\V{k} \in ]0,2 k_0[$.
This translate to a characteristic (or integral) length scale of
$L = 2$. 

In \textsc{Enzo} we use a stochastic acceleration field that evolves in 
space and time as an Ornstein–Uhlenbeck process\citep{Schmidt2009}. 
The forcing modes are projected from spectral to real space so
that they are split into 50\% compressive and 50\% solenoidal
modes.
In \textsc{Athena} we use a random acceleration field that is purely
solenoidal.
It is regenerated every $0.2T$ large-eddy turnover times.

In general, we follow the simulations for a total of $5T$ 
turnover times with $T = L/V$.
The characteristic velocity $V$ is given by the r.m.s.~Mach number
in the stationary regime, i.e. $0.5$ in the subsonic case and $2.5$ in the
supersonic case.
We disregard data during the initial $2T$ where turbulence develops from
the uniform initial conditions.
During the stationary regime $2T \leq t \leq 5T$ we captures 31 snapshots
equally spaced in time.
If not otherwise noted, all quantities in the following analysis are 
given by the mean values over these 31 snapshots and variations given by 
one standard deviations.

\section{Results}
\label{sec:results}

The results presented here 
highlight the uses of the analysis technique.
In order to get a comprehensive picture of the energy transfers,
we start from an aggregated point of view and then go into more and
more detail.
First, we compare general flow properties and energy spectra as a basis
for the discussion.
Second, we analyze the energy fluxes across scales.
Third, we describe the total transfer, i.e. how much energy is gained
(or lost) at a particular scale. 
Finally, we describe the individual shell-to-shell transfers.

\subsection{Flow properties and energy spectra}
\label{sec:ResFlow}
\begin{table*}[!htbp]
        \begin{tabular}{lcccccccc}
                \toprule
                Run  &  $\langle \Ekin \rangle$ & $\langle \Emag \rangle$ &
                $\langle \Emag \rangle / \langle \Ekin \rangle$&                
                $\langle \pb \rangle$&                
                $\langle \langle \mathrm{M}_{\mathrm{s}}^2 \rangle^{1/2} \rangle$ & 
                $\langle \langle \mathrm{M}_{\mathrm{a}}^2 \rangle^{1/2} \rangle$ &                  
                {$\left < \Pi \right >$}
                \\
\colrule
$\mathtt{M0.5-Enzo}$	& 0.16 & 0.21 & 1.36 & 23.09	& 0.57 & 1.81 &{0.352}\\
$\mathtt{M0.5-Athena}$	& 0.13 & 0.16 & 1.22 & 27.35	& 0.51 & 1.89 & {0.246}\\
$\mathtt{M2.5-Enzo}$	& 2.88 & 2.39 & 0.83 & 1.34	& 2.51 & 2.11 & {24.3}\\
$\mathtt{M2.5-Athena}$	& 3.30 & 2.78 & 0.84 & 1.09	& 2.63 & 1.95 & {37.7}\\
\botrule
        \end{tabular}
        \caption{Overview of the flow properties,
		i.e. mean kinetic energy $\Ekin$,
		mean magnetic energy $\Emag$, ratio of
		mean magnetic to kinetc energy,
		mean plasma $\pb$, rms sonic
		Mach number $\Ms$ and rms 
		Alfv\`enic Mach number $\Ma$,
		in the simulations
		during the stationary phase.
    {The last column lists the mean (over time) of the mean total fluxes
    (in the inertial range), which are used for normalization in manuscript.}
        \label{tab:SimOverviewFlow}} 
\end{table*}
Table~\ref{tab:SimOverviewFlow} shows mean and root mean square (rms) characteristic
flow quantities during in the stationary regime.
The two simulations in each regime are very similar independent of the numerical 
method.
In the subsonic regime, the rms sonic Mach number is $\approx0.5$ and the
mean plasma $\pb$ is $\approx25$.
In the supersonic regime, the sonic Mach number is $\approx2.5$ and 
$\pb$ of order unity.
All simulations are in the super Alf\'enic regime with rms Alf\'en Mach number
$\approx2$.
While there is more energy in the magnetic reservoir than in the kinetic reservoir
in the subsonic regime ($\langle \Emag \rangle /\langle \Ekin \rangle \approx 1.3$),
the opposite is true in the supersonic regime where the ratio is $\approx 0.83$.

The larger ratio in the subsonic regime is also present on all scales as
shown in the bottom panel of Fig.~\ref{fig:Specs} where the 
ratio is plotted over wavenumber.
In the subsonic regime magnetic energy is dominant on all scales beyond
the forcing range.
In the supersonic regime it is only dominant beyond the large scale
forcing and up to $k\approx30$.
This range coincides with the range where approximately power-law scaling
is observed in the kinetic energy spetrum, as can be seen in the top panel of 
Fig.~\ref{fig:Specs}.
All simulations have a flat kinetic energy spectrum 
between $7 \lesssim k \lesssim 30$ when compensated by $k^{4/3}$.
For this reason, we use this region as the ``inertial range'' in the following
analysis.
The compensated (by $k^{1.7}$) magnetic energy spectra in the center 
panel lack any clear power-law regime, a result that is independent of regime and numerical
method and in agreement with previous studies\cite{Kritsuk2011,Porter2015}.

Overall the spectra are again very similar within each regime and
differences due to the numerical method become first apparent at the smallest
scales.
For example, the supersonic Enzo simulation is the only simulation that uses
the HLL Riemann solver, whereas the other three runs use the less dissipative 
HLLD Riemann solver.
This translates to a more pronounced decay of the spectrum at high wavenumbers
and a slight decrease of the intertial range.
Similarly, differences close to the grid scale are visible in all simulations,
especially in the ratio of magnetic to kinetic energy.
This is no surprise as this region is dominated by the numerical method itself.
For the purpose of the manuscript and our analysis this is of no importance
as contributions from these scales are negligible.

\subsection{Cross-scale energy fluxes}
\label{sec:ResCrossScale}
\begin{figure*}[!htbp]
\centering
\includegraphics{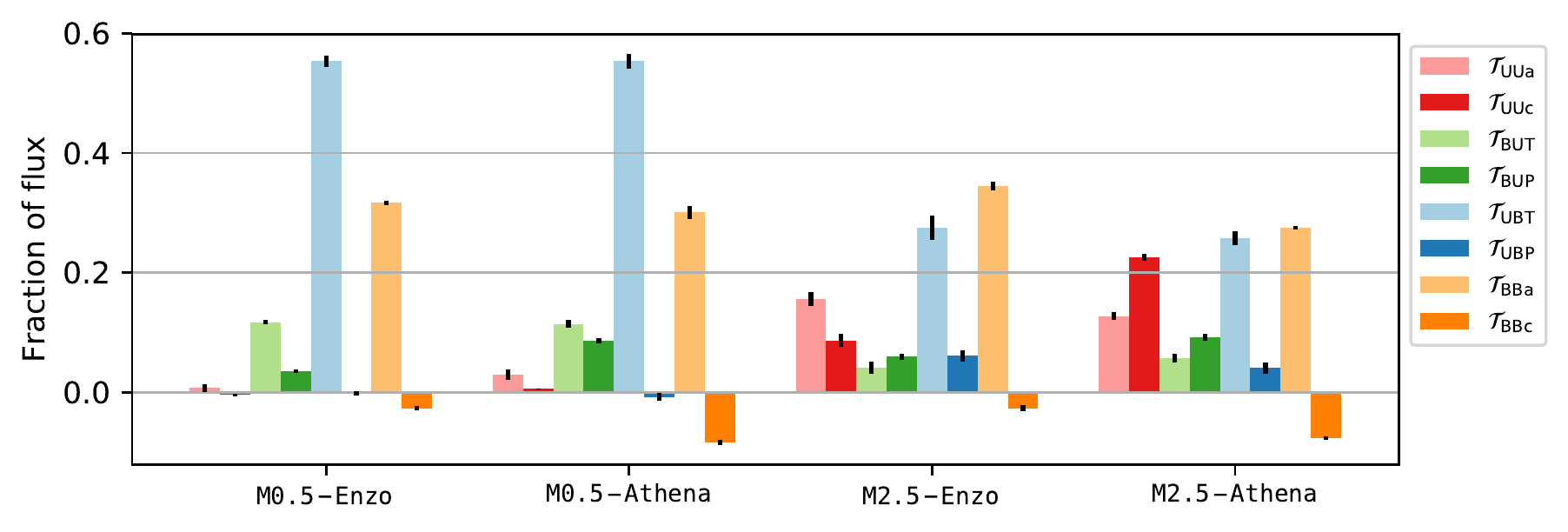}
\caption{
Mean fraction of the cross-scale flux in the 
inertial range by the individual terms in the stationary phase.
The black lines at the end of each block indicate the standard
deviation over time.
The fluxes have been normalized to sum up to $1$.
}
\label{fig:InertialFlux}
\end{figure*}
We begin the energy transfer analysis from a broad point of view:
the mean cross-scale energy flux within the inertial range as illustrated
in Fig.~\ref{fig:InertialFlux}.
We follow the normalization of Debliquy \textit{et al.}\cite{Debliquy2005}.
Thus, each snapshot is normalized to the mean
total flux in the inertial range
{
  \begin{align}
  \begin{split}
\left < \Pi \right > = \sum_{k \in [7,30]} & \Big( 
\Pi^{\mathrm{U}^<}_{\mathrm{U}^>} (k) + 
\Pi^{\mathrm{U}^<}_{\mathrm{B}^>} (k) + 
\Pi^{\mathrm{B}^<}_{\mathrm{U}^>} (k) + \\ &
\Pi^{\mathrm{B}^<}_{\mathrm{B}^>} (k) + 
\Pi^{\mathrm{P}^<}_{\mathrm{U}^>} (k) + 
\Pi^{\mathrm{F}^<}_{\mathrm{U}^>} (k)  \Big) \;.
\end{split}
\end{align}
}
We use this normalization {for all quantities throughout the manuscript, i.e.
all transfer functions for a particular snapshot are divided by the respective
$\left < \Pi \right >$ of that snapshot.}
Overall there is little variability with respect to time.
In the subsonic
regime the cross-scale flux is dominated by kinetic to magnetic transfer
by magnetic tension (contributing approximately $55\%$) 
and magnetic to magnetic transfer by advection ($\approx 31\%$).
The contribution of the kinetic cascade is negligible and there is only a limited
energy flux from magnetic energy at large scale to smaller scales by 
magnetic tension ($\approx11\%$) and magnetic pressure ($\approx 4\%$
for \textsc{Enzo} and $\approx8\%$ for \textsc{Athena}).
It should be noted that these fractions are very close the results
obtained by Debliquy \textit{et al.} (see Table~I in 
Ref.~\onlinecite{Debliquy2005}).
For example, they report fluxes by magnetic tension of $49\%$ from the 
kinetic to the magnetic budget and of $12\%$ from the magnetic to the kinetic budget.
In addition, their kinetic flux by advection is similarly small ($7.5\%$) and 
the magnetic flux by advection contributues with $37\%$.
Contrary to the compressible finite volume codes we use, they
employed a fully-dealiased pseudospectral code and analyzed
decaying incompressible MHD turbulence at a resolution of $512^3$.

The differences between the subsonic and the supersonic regime are
 visible upon first inspection.
While the magnetic cascade fluxes are unchanged, the contribution
of kinetic to magnetic cross-scale flux by magnetic tension is cut
in half to about $26\%$.
This is generally compensated by the kinetic cascade fluxes 
even though they vary slightly between \textsc{Enzo} (with 
$15\%$ by advection and $8.4\%$ by compression) and \textsc{Athena}
with $12.7\%$ and $22.4\%$, respectively).
Moreover, magnetic pressure now contributes with $\approx 7\%$ to 
a large scale magnetic to small scale kinetic energy flux, which 
was absent in the subsonic regime.
Overall the picture in the supersonic regime is much more balanced
between the individual mediators and there is no single dominant
contribution.

One interesting feature in the mean cross-scale fluxes is the magnetic
cascade term mediated by compressive effects.
One the one hand it is consistently negative, i.e. there is a magnetic
energy flux from small to large scales, in all simulations.
In the other hand, its seems to be more affected by the numerical
method rather than the regime.
For \textsc{Enzo} it contributes about $-2.5\%$ in both regimes,
whereas for \textsc{Athena} it is more pronounced with approximately
$-8\%$.

\begin{figure*}[!htbp]
\centering
\includegraphics{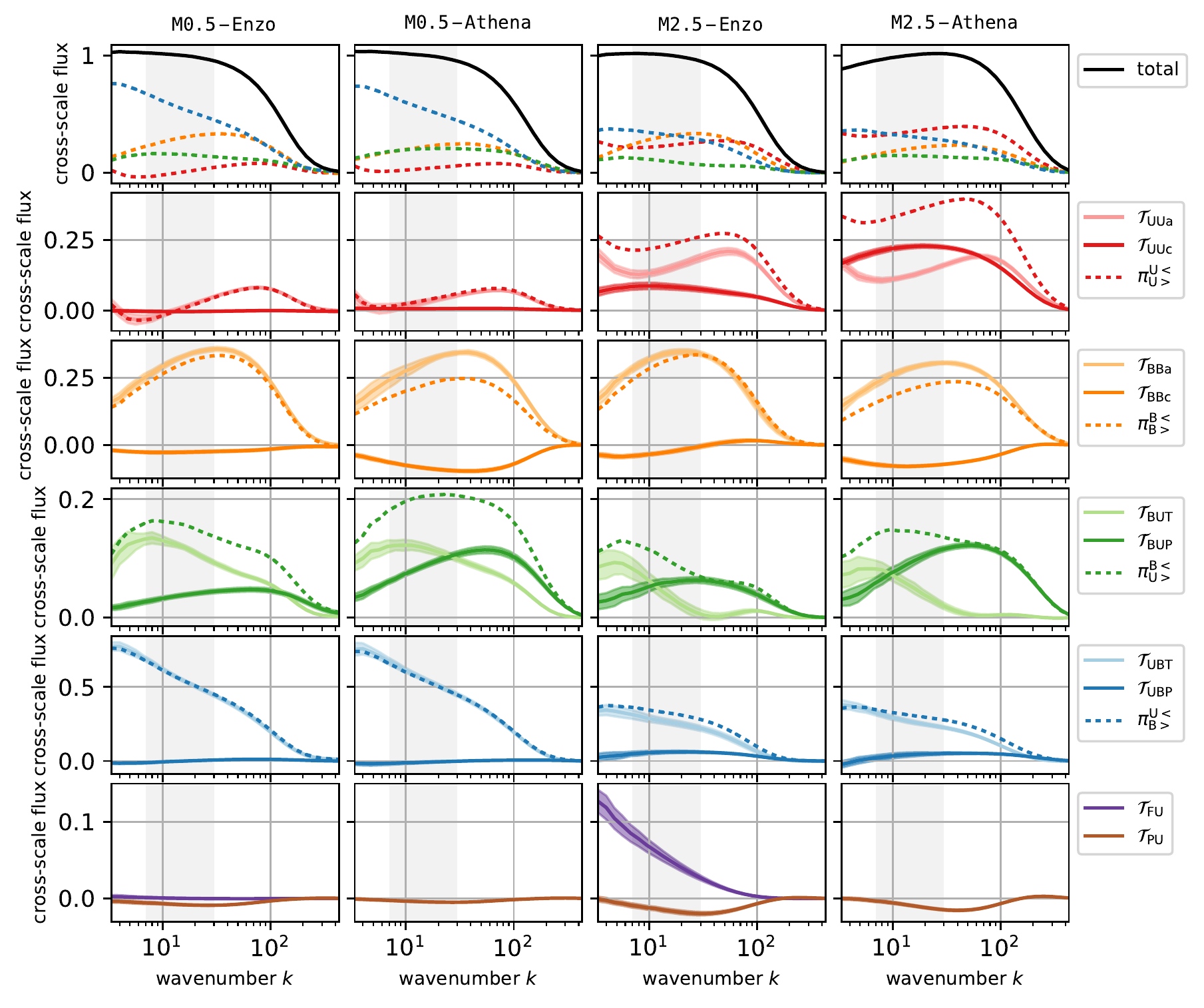}
\caption{
	Mean cross-scale fluxes versus wavenumber of the individual
	terms during the stationary phase.
    Variations by means of the standard deviation are shown by
	the vertical shaded regions.
    The first row includes the joint fluxes between kinetic and
    magnetic energy reservoirs
    and the total cross-scale flux for comparison.
	The total cross-scale flux from large scale kinetic energy to
	kinetic energy on smaller scales 
	($\Pi^{\mathrm{U}^<}_{\mathrm{U}^>}$, second row), large scale magnetic
	to small scale magnetic 
	($\Pi^{\mathrm{B}^<}_{\mathrm{B}^>}$, third row), large scale magnetic to
	small scale kinetic 
	($\Pi^{\mathrm{B}^<}_{\mathrm{U}^>}$, fourth row) and large scale kinetic to
	small scale magnetic 
	($\Pi^{\mathrm{U}^<}_{\mathrm{B}^>}$, fifth row) are highlighted by a dashed line.
	For completeness, the bottom row shows the cross-scale flux by pressure variations
	and by the external forcing (data only available in \textsc{Enzo}). 
\label{fig:CrossFlux}
}
\end{figure*}

A more detailed picture of the cross-scale fluxes is shown 
in Fig.~\ref{fig:CrossFlux}, where the fluxes are now plotted
over all wavenumbers.
For all simulations the general shape of the individual fluxes
is similar. It is predominately the scale that varies between regimes.
It is important to note that the joint total fluxes by all terms 
together are approximately constant (black lines in the top row) until 
the end of the inertial range (highlighted by the gray areas).
This is expected from theory for an inertial range {in 
the incompressible regime}.
However, the individual fluxes are not at all constant 
and change substantially with $k$.
For example, the dominant contribution of $\T{UBT}$ in the subsonic
regime is constantly decreasing within the inertial range whereas
the contribution by $\T{BBa}$ is constantly increasing.

Overall the contribution of $\T{BBa}$ is very robust with respect
to shape and magnitude as illustrated in row 3 of Fig.~\ref{fig:CrossFlux}.
Its contribution is doubling from $\approx 16\%$ at the largest scales
to $\approx 32\%$ at the end of the inertial range where it is then 
damped by numerical dissipation.
The previously observed mean inverse flux in the magnetic cascade term via
compression $\T{BBc}$ in the inertial range now clearly extends beyond 
the wavenumbers $7 < k < 30$.
It is also more pronounced in the \textsc{Athena} simulations independent
of the regime.

This feature is similarly present in the kinetic cascade term via compression
$\T{UUc}$, which is, albeit having the same shape, 2.5-3 times as strong in
\texttt{M2.5-Athena} than it is in \texttt{M2.5-Enzo}.
As expected this term is practically absent on all scales in the subsonic regime
and independent of numerical method.
In general, the cross-scale fluxes from kinetic energy at large scales 
($\Pi^{\mathrm{U}^<}_{\mathrm{U}^>}$ and $\Pi^{\mathrm{U}^<}_{\mathrm{B}^>}$)
exhibit the largest quantitative changes going from the subsonic to 
the supersonic regime.

Another interesting feature concerns the large scale magnetic to small
scale kinetic cross-scale energy transfer 
$\Pi^{\mathrm{B}^<}_{\mathrm{U}^>}$ and its components $\T{BUT}$ and $\T{BUP}$ (see row 4 of Fig.~\ref{fig:CrossFlux}).
While in the subsonic regime the magnetic tension related flux is consistently
non zero up to the smallest scales in the domain, it is effectively
zero beyond the inertial range in the supersonic regime.
This leads to interesting dynamics in combination with the flux by magnetic pressure.
Given that latter is always non zero and positive, the dominant contribution 
to $\Pi^{\mathrm{B}^<}_{\mathrm{U}^>}$  changes with scale and in the
supersonic regime changes within the inertial range from tension dominated 
to pressure dominated.

Finally, the direct cross-scale transfer by the external mechanical driving
is of importance.
Independently of defining an acceleration field, as done here, see \eqref{eq:FU},
or defining a forcing field, the external driving is always coupled to
density field.
Thus, our acceleration field, which is confined to $k \in ]0,4[ $, actually
injects energy beyond $k = 4$.
This is most obvious in the bottom row of Fig.~\ref{fig:CrossFlux} where
the cross-scale fluxes for the \textsc{Enzo} simulations are plotted.
In the subsonic regime it is practically zero on all scales whereas in the
supersonic regime it constantly decreases from the largest scales down
to $k\approx100$.
It is directly linked to the different extent of density fluctuations between
the sub- and the supersonic regime.
We only have data available for the \textsc{Enzo} simulation.
The absence of that term is clearly visible for the total cross-scale flux 
in \texttt{M2.5-Athena}, which is not constant (in our analysis) 
on the largest scales for that reason.

\subsection{Total energy transfer}
\label{sec:ResTotal}

\begin{figure*}[!htbp]
\centering
\includegraphics{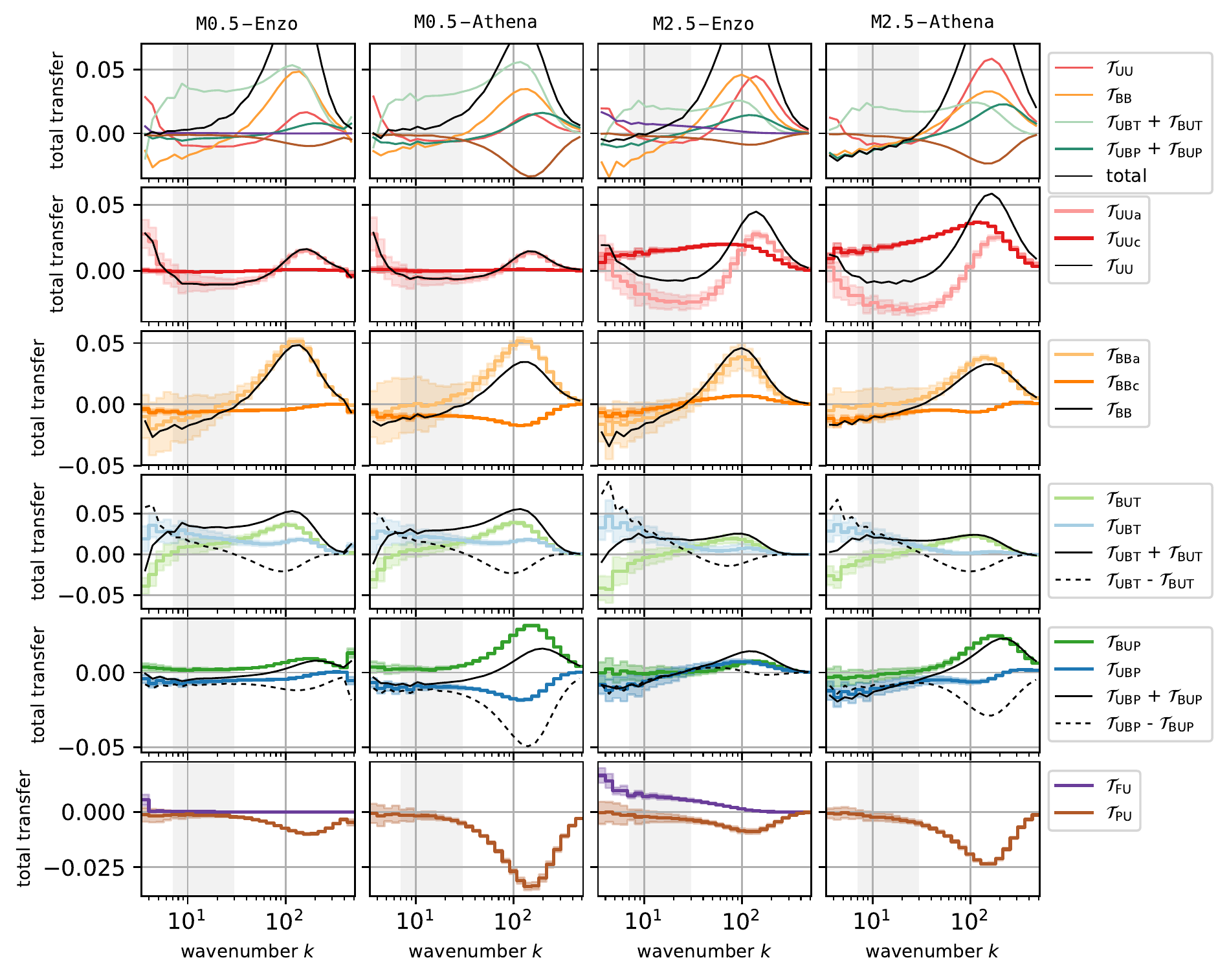}
\caption{
	Mean total transfer in (positive) or out (negative) of
	a shell from (or to) all other shells.
	The transparent regions show the standard deviation over time.
    For comparison, all transfers including the total transfer are
    illustrated in the first row.
	The second row shows transfer from the kinetic energy reservoir 
	to kinetic energy on other scales, 
	the third row magnetic to magnetic transfer,
	the fourth row magnetic to kinetic transfer and the
	fifth row kinetic to magnetic transfer.
	The bottom row shows the transfers that are not within or between
	kinetic and magnetic energies, i.e. transfer by the external force (data
	only available for \textsc{Enzo}) and pressure.
	The dashed lines in the magnetic tension and pressure related panels
	illustrate the net transfer to magnetic energy.
  {All transfer functions follow the same normalization as before, i.e.
  they are normalized by the total mean inertial cross-scale flux 
  $\left< \Pi \right>$.}
	}
\label{fig:TotalTransfer}
\end{figure*}

After the presentation of the energy transfer across scales, we now
present the total energy transfer into and out of particular scales, i.e.
$\sum_Q\T{} (Q,K)$.
Figure~\ref{fig:TotalTransfer} illustrates these transfers for all terms over
all wavenumbers beyond the direct forcing regime.
As with the cross-scale fluxes the overall picture is qualitatively very similar across
the simulations, but has some notable quantitative differences.

The kinetic cascade term $\T{UU}$ is approximately zero within the
inertial range.
However, there are significant differences between the subsonic and  
supersonic regimes.
While in the former regime both contributions ($\T{UUa}$ and $\T{UUc}$) are 
effectively zero, they are negative (advective term) and positive 
(compressive term) in the supersonic regime.
Thus, compression and advection work against each other and their contributions
cancel.

Similarly, the magnetic cascade term is approximately zero within the 
inertial range, but the individual terms are practically identical
in both regimes.
The advective component $\T{BBa}$ exhibits large variations around zero 
throughout the inertial range whereas the compressive component $\T{BBc}$
is constantly negative with only little variability.

The magnetic tension related terms ($\T{BUT}$ and $\T{UBT}$) provide an 
interesting insight into the
dynamics between kinetic and magnetic energies.
On the one hand, their joint effect leads to an increase of the total energy 
on all scales (see solid black line in the fourth row of 
Fig.~\ref{fig:TotalTransfer}).
On the other hand, that increase of total energy is in magnetic energy on large
and intermediate scales (positive dashed black line) and in kinetic energy
on smaller scales (negative dashed black line).
The zero crossing takes place at the end of the inertial range in all 
simulations.
Whether this occurs by coincidence or due to the underlying physics should
be verified in additional simulations at different resolutions.
Moreover, this profile is also quantitatively fairly independent of the regime
and numerical method.

Transfers mediated by magnetic pressure show a less consistent picture between 
individual runs, see fifth row of Fig.~\ref{fig:TotalTransfer}.
In the supersonic regime the variations in both terms throughout the inertial
range are consistent with zero. 
In the subsonic regime there is less variability and the kinetic to magnetic
transfer $\T{UBP}$ is clearly negative and slight more pronounced in the
\textsc{Athena} simulation.
Interestingly, magnetic pressure is thus working against magnetic tension
in transferring energy from the kinetic reservoir to the magnetic 
one within the inertial range as $\T{UBT}$ is positive over these wavenumbers.

As already seen in the cross-scale transfers, the forcing term is non-zero
beyond the direct forcing scales (bottom row of Fig.~\ref{fig:TotalTransfer})
in the supersonic regime.
Even though its overall contribution is weak, some energy is directly injected
to the kinetic reservoir on intermediate scales.

Finally, two additional features should be noted.
First, the profile of the total transfer of all terms (black line in the
top row of Fig.~\ref{fig:TotalTransfer}) starts to deviate from zero
beyond the inertial range.
This is a direct measure of the numerical dissipation of the individual
numerical methods.
Second, from a total transfer point of view $\T{BBc}$ and
$\T{UBP}$ are almost identical.
Similarly, $\T{BBa}$ and $\T{BUP}$ exhibit the same dynamics.
This can be related to their derivation, see \eqref{eq:MagTermMagPres} 
and \eqref{eq:TUBPwBB}.
However, this impression is misleading as their underlying shell-to-shell
transfer is  different as shown in the following subsection.

\subsection{Shell-to-shell transfer}
\label{sec:ResShell}

\begin{figure*}[!htbp]
\centering
\includegraphics{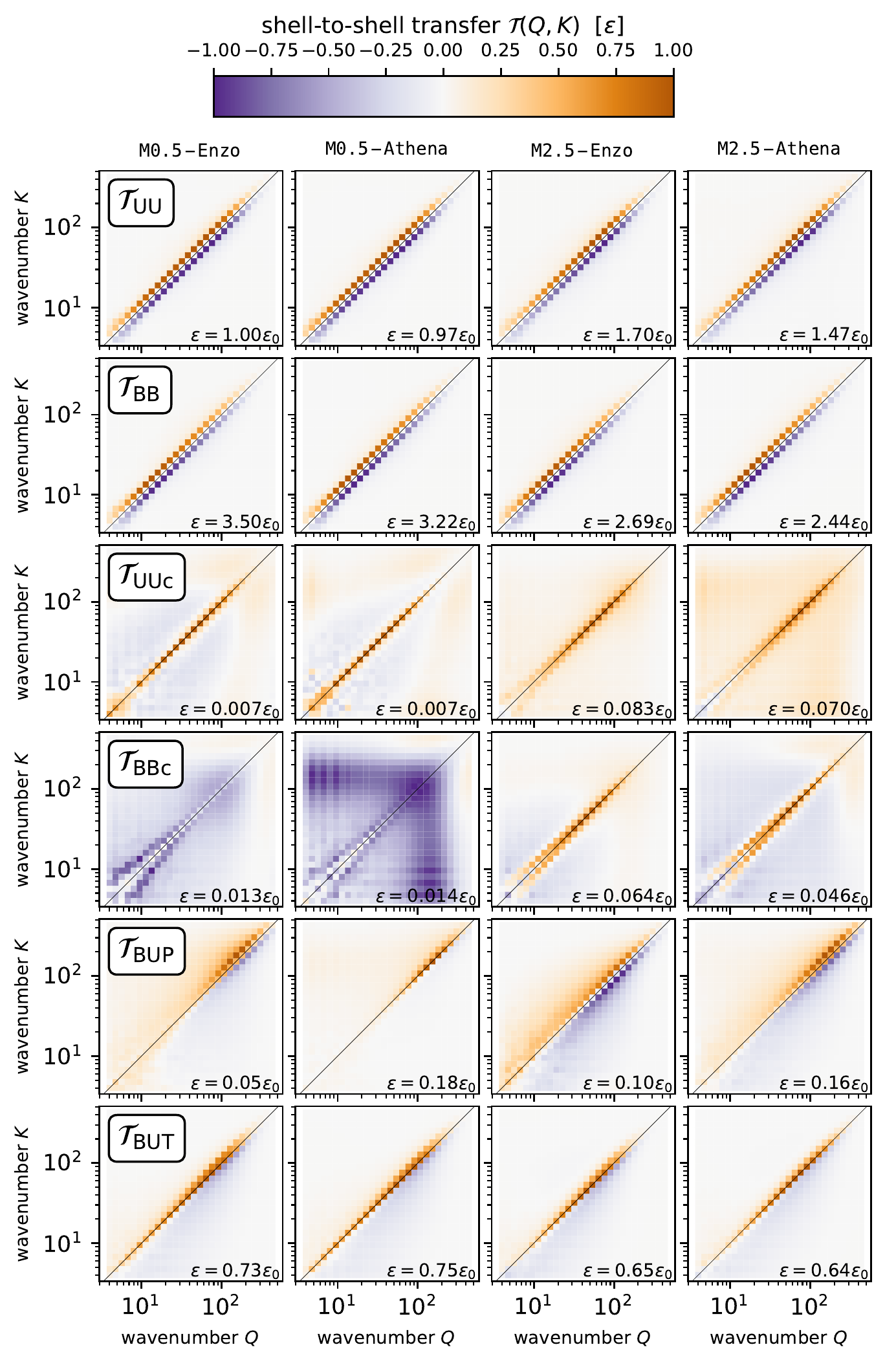}
\caption{Mean transfer from shell $Q$ to shell $K$ by the individual 
transfer terms (from top to bottom):
$\T{UU}$, $\T{BB}$, $\T{UUc}$, $\T{BBc}$, $\T{BUP}$ and $\T{BUT}$.
The transfers for $\T{UBP}$ and $\T{UBT}$ are not explicitly shown.
Due to the antisymmetry property their illustrations are are identical to 
the transposed panels of the shown $\T{BUP}$ and $\T{BUT}$ terms,
respectively.
Each panel uses a unique symmetric colorbar that extends from $-\varepsilon$ to
$\varepsilon$.
All are normalized to the maximum $\T{UU}$ value of simulation \texttt{M0.5-Enzo}
with $\varepsilon_0 = 0.052$.
\label{fig:ShellTransfer}
}
\end{figure*}

We close the analysis section with the presentation of the individual 
shell-to-shell transfers for all terms as depicted in 
Fig.~\ref{fig:ShellTransfer}.

In agreement with earlier work, we find the kinetic $\T{UU}$ and magnetic 
$\T{BB}$ cascade transfer function to be highly local, as shown in panels
in the top two rows of Fig.~\ref{fig:ShellTransfer}.
Independent of regime and numerical method, energy is received from the
next shell on larger scales and released to the next shell on smaller scales.
While in the subsonic regime the magnetic cascade is more than
three times as strong as the kinetic cascade, it is a little more balanced
in the supersonic regime.
Here, the kinetic cascade is $\approx 1.6$ stronger compared to the 
\texttt{M0.5} runs and the magnetic cascade is less than twice as 
strong as its kinetic counterpart
as indicated by the different scaling coefficient $\epsilon$.
In general, each panel is normalized with an individual $\epsilon$,
see caption of Fig.~\ref{fig:ShellTransfer}.

The compressive $\T{XXc}$ components
of $\T{UU}$ and $\T{BB}$ are plotted in rows three and four of 
Fig.~\ref{fig:ShellTransfer}.
The kinetic compressive component is overall quite consistent between
numerical methods.
However, going from the \texttt{M0.5} to \texttt{M2.5} increases its 
strength by a factor of $10$ even though it is overall still weak compared
to the full cascade term $\T{UU}$.
The full term is zero on the diagonal, i.e. $K = Q$ by construction.
This is not true of its components and $\T{UUc}$ is actually strongest
(and positive) on the diagonal.
Conversely, $\T{UUa}$ (not shown) must be negative on the diagonal illustrating
how advection and compression work against each other as already mentioned.

The compressive magnetic cascade term $\T{BBc}$ is in stark contrast to its
kinetic counterpart.
The regime plays an important role.
For both subsonic runs the transfer is entirely negative independent of
the giving and receiving wavenumber.
Moreover, the numerical method also introduces differences.
In \texttt{M0.5-Enzo} the transfer predominately takes place off the diagonal and
between large scales whereas in \texttt{M0.5-Athena} the strongest transfers 
are on the small scales.
In the supersonic regime the picture is more consistent with respect to codes.
Contrary to the subsonic regime, energy is now received from shells on and adjacent
to a particular shell $K$, and transfers that are not close to the diagonal are 
comparatively weak.

As mentioned in the previous subsection, magnetic pressure and magnetic cascade
transfers seem to be almost identical from a total transfer point of view.
However, the individual shell-to-shell analysis 
(see fifth row of Fig.~\ref{fig:ShellTransfer}) reveals that $\T{BUP}$ is 
much less localized in spectral space.
Energy is now transferred not only between the next larger and smaller shell, but 
also between the next 3-4 shells in both directions.
In addition, a difference in the numerical methods is seen in the (numerically) 
dissipative regime.
Independent of regime, there is no significant transfer of energy between kinetic
and magnetic energies at the same scale ($K=Q$) for \textsc{Enzo}.
In \textsc{Athena} a non negligible amount of energy is transfered from magnetic
to kinetic energy at the same scale.
This transfer is overall the strongest transfer mediated by magnetic pressure in the
subsonic regime.

Finally, the shell-to-shell transfer via magnetic tension is in agreement 
with previous results\citep{Aluie2010,Teaca2011}.
It is weakly local and also most pronounced on the diagonal, i.e.
energy is directly transferred between magnetic and kinetic reservoirs at
the same scale.
Moreover, this feature is consistent between regimes (with less than $20\%$
difference in strength) and independent of numerical method.

\section{Discussion}
\label{sec:discussion}
The energy dynamics presented in Section \ref{sec:results} provide an
interesting insight into both compressible MHD turbulence
dynamics and numerics.
For example, our results in the subsonic regime are in  good
agreement with previous studies conducted in the incompressible
MHD regime\cite{Debliquy2005,Aluie2010,Teaca2011}.
This agreement is even more relevant given that these studies
employed a spectral code whereas we employ fully compressible
finite volume codes.
However, subtle differences between the two methods we used
are revealed by the analysis, e.g., for the inverse fluxes of
the compressible magnetic cascade term on the smallest scales.
A more detailed study of the presented energy dynamics in 
identical setups with different numerical methods including spectral and finite 
difference methods would be informative.
Such an analysis would allow for the quantification of the practical influence
of the numerical scheme on the dynamics and a measurement
of numerical dissipation similar to Salvesen \textit{at al.}\cite{Salvesen2014}.

The observed upscale flux in the compressible magnetic cascade term is 
also relevant for subgrid-scale models in large eddy 
simulations\cite{Miesch2015}.
Purely dissipative models, such as an eddy viscosity or resistivity,
are prone to overestimate the downscale flux.
Structural models, e.g. the nonlinear model by 
Vlaykov \textit{et al.}\cite{Vlaykov2016a} that explicitly captures 
compressibility effects in MHD, are capable of reproducing inverse 
fluxes\cite{Grete2016a}.
The local nature of the energy transfer in the kinetic and magnetic
cascade terms seems to be ideally suited for scale similarity-type 
subgrid-scale models\cite{BARDINA1980}.
However,  Grete \textit{et al.} (2017)\cite{Grete2017}
showed that contrary to the nonlinear model, the scale similarity model did
not improve higher order statistics
when applied to decaying, supersonic MHD turbulence.

From an energy transfer point of view, our extension of the
shell-to-shell analysis method to the compressible regime allows for a
separation of a magnetic cascade and transfer by magnetic pressure,
see \ref{sec:DefBBandBUP}.
On the one hand some previous studies claimed\cite{Graham2010,Moll_2011}
that magnetic energy transfer between 
different scales (the magnetic cascade) 
and the exchange of energy between kinetic and magnetic 
reservoirs via magnetic pressure cannot be separated.
On the other hand this separation has already been shown
in the context of total transfer functions\cite{Simon2009}.
Here, we go one step further and split the total transfer functions into
shell-to-shell transfer functions.
This {view} allowed us to illustrate that the magnetic cascade is quite different from the 
magnetic pressure related terms at a shell-to-shell level.
{For example, $\T{BBc}$ and $\T{UBP}$ are identical at the level of 
total energy transfers.
However, from an individual shell-to-shell point of view they are very dissimilar.}

Another important result revealed by the analysis concerns
the external force used in driven
turbulence simulations.
The forcing (or acceleration) field is inescapably coupled 
to the density field.
This is of no concern in the subsonic regime due to limited 
density fluctuations.
While the supersonic regime in our simulations is only mildly 
supersonic with $\Ms \approx 2.5$, we already observe a non-negligible 
energy input throughout the inertial range.
{On the one hand, } this effect, i.e. energy injection beyond the range of scales defined in the
external field, is expected to be much more pronounced in the highly supersonic
regime.
{On the other hand, Aluie\cite{Aluie2013} proved in the context of a Favre filtering based
scale decomposition that transfers via the forcing term vanish for $K \gg Q$.
This is also observed in our approach, see Fig.~\ref{fig:TotalTransfer}.
Overall,} analyzing driven turbulence data in {highly supersonic} regimes (e.g., with respect to
slopes of the spectral energy densities) should thus be done with special care,
{especially in presence of a limited inertial range}.

Moreover, Domaradzki \textit{et al.}\cite{Domaradzki2010} showed that even in the incompressible
MHD regime 
{the directly forced scales implicitly affect the transfer functions.
Even though there is no direct energy injection by the $\T{FU}$ function beyond
the forced scales, the nature of triadic interactions allow for the mediator
to lie within the forced scales.
The latter typically contain the most power and carry an imprint of the external
forcing field.
Thus, individual triadic interactions between (typically weaker) small scales are
affected by a leg in the forced scales.}
It is a natural effect occurring in all forced turbulence simulations.
{In this manuscript, we are not concerned with individual interactions but use
mediators that  contain information of all wave numbers, which always
include interactions with the forcing scales.
The same approch is used by Domaradzki \textit{et al.}\cite{Domaradzki2010} and } they 
estimate a minimum resolution well above $1000^3$ in order to see a decoupling
of the forcing scales from the smallest resolved scales.
While the resolution of our simulations is close to that number, we expect
it to be greater in the case of compressible MHD (with shock-capturing 
finite volume methods) --- especially if more than the smallest resolved
scales should be decoupled.

In addition to the forcing, several other 
characteristics of our present analysis should be kept in mind.
For example, the third quantity in the transfer functions is not
restricted to a particular scale.
As noticed by Alexakis \textit{et al.}\cite{Alexakis2005}, 
this translates to all transfer 
functions containing information only about whether the energy transfer 
itself is local and not about
whether the interaction itself is local.
However, as pointed out by Aluie \& Eyink\cite{Aluie2010} the choice of the shell
spacing directly influences whether local or nonlocal interaction are favored.
For the logarithmic binning we use the number of possible local interactions
is much larger than the number of possible nonlocal interaction. 
Thus, our results may also be interpreted as a hint about the locality of the 
interactions, but a more detailed study is required. 
Similarly, the present study is concerned with one of the simpler MHD turbulence
configurations (super-Alfv\'enic, isotropic, non helical) as a proof of concept
for the analysis.
Again follow-up studies in different regimes, such as turbulence with a strong
background magnetic field or in the presence of an inverse magnetic helicity
cascade, are desirable.

\section{Conclusions}
\label{sec:conclusions}
In this paper we extended an established shell-to-shell energy transfer 
analysis to the compressible MHD regime.
Four transfer functions regulating energy transfer within and between kinetic 
and magnetic reservoirs are known from the incompressible formalism:
kinetic to kinetic transfer by advection (the kinetic cascade), 
magnetic to magnetic transfer (the magnetic cascade), and transfer
from kinetic to magnetic energy (and vice versa) by magnetic tension.
We derived four additional terms are present in the compressible 
formalism that can be separated into two categories.
First, both known cascade functions now contain not only the advective term
but also a second term directly associated with the compressibility 
$\nabla \cdot \V{u}$ .
Second, an additional channel to exchange energy between kinetic and magnetic 
reservoirs is possible.
{We interpret} this channel {as} mediated by magnetic pressure, which 
adds two (antisymmetric) transfer functions.

In order to illustrate the value of this formalism, we conducted and
analyzed four simulations of driven, isothermal, ideal MHD turbulence.
Using two different numerical codes, \textsc{Enzo} and \textsc{Athena}, 
in two regimes, subsonic ($\Ms \approx 0.5$) and supersonic 
($\Ms \approx 2.5$), we were able to identify similarities and differences
both with respect to MHD turbulence and numerical method.

In addition to the pure shell-to-shell transfer we also analyzed 
higher-level transfers, e.g. energy transfer across scale and the total
transfer into (or out of) specific scales by the individual functions.
Differences between regimes became apparent.
While the cross-scale fluxes in the subsonic regime are dominated by 
kinetic to magnetic energy transfer via tension and magnetic to magnetic 
transfer via advection,  the supersonic regime provides a more diverse
impression.
In this regime, kinetic to kinetic transfer by both advection and compression and 
kinetic to magnetic transfer via magnetic pressure contribute
with non-negligible amounts.
From a total transfer point of view the net effect of the 
kinetic cascade is similar in both regimes.
Despite similarity in the net effect, while  in the subsonic regime the compressive component is effectively
absent in the supersonic regime both advective and compressive transfer are much 
stronger but also work against each other.
The shell-to-shell results are also in agreement with previous findings.
Kinetic and magnetic energy both exhibit features of a forward local energy 
cascade.
Nonetheless, it is important to mention that this only holds for the functions 
that contain the advection and the compressive component.
As known from the incompressible regime magnetic tension-related functions 
are weakly local and forward.
Our analysis shows that this is also true for the new magnetic pressure-related
functions.

Overall the application of shell-to-shell energy transfer analysis 
exposed several interesting features that would benefit from more detailed
follow-up studies.
For example, the dynamics of the compressive component in the magnetic
cascade term concerning cross-scale fluxes seem to be dominated by numerical 
method rather than physical regime.
Then again, the dynamics of that term concerning shell-to-shell transfer are
quite similar between methods but differ a lot between regimes.
Another example is the magnetic to kinetic cross-scale flux.
It is dominated by magnetic tension on the large scales and dominated by
magnetic pressure on the small scales.
There are indications for a trend in characteristic features, e.g. the scale 
where both are equally strong or where magnetic tension becomes 
completely negligible, but more data (with different parameters) is required
to draw definite conclusions.
This also applies to the net total transfer by magnetic tension.
In all simulations independent of regime and  method the scale where
predominately kinetic to magnetic transfer changes to predominately magnetic
to kinetic transfer coincides with the end of the inertial range.
Here, higher resolution simulations would allow one to trace that behavior
in the presence of an extended inertial range.

\begin{acknowledgments}
The authors thank David Collins, Sean Couch, Jeffrey Oishi, Dimitar Vlaykov
and Dominik Schleicher for useful discussions.
PG, BWO, KB and AC acknowledge funding by NASA Astrophysics Theory Program 
grant \#NNX15AP39G, KB acknowledge funding by NASA Astrophysics Theory Program 
grant \#NNX14AB42G and BWO acknowledges additional funding by NSF AAG grant \#1514700.
This work was performed (in part) under the auspices of the Air Force Office of Scientific Research Work Performed under Contract \#FA9550-17-C-0010.
The \textsc{Enzo} simulations were originally conducted with
the HLRN-III facilities of the North-German Supercomputing 
Alliance under Grant nip00037.
The \textsc{Athena} simulations were run on the NASA Pleiades supercomputer through allocation SMD-16-7720.
This research is part of the Blue Waters sustained-petascale computing project, which is supported by the National Science Foundation (awards OCI-0725070 and ACI-1238993) and the state of Illinois. Blue Waters is a joint effort of the University of Illinois at Urbana-Champaign and its National Center for Supercomputing Applications.
This work is also part of the PRAC ``Petascale Adaptive Mesh Simulations of Milky Way-type Galaxies and Their Environments" PRAC allocation support by the National Science Foundation (award OCI \#1514580).
This work was supported in part by Michigan State University through computational resources provided by the Institute for Cyber-Enabled Research.  \texttt{Enzo}
and \texttt{Athena} are developed by a large number of independent
researchers from numerous institutions around the world. Their
commitment to open science has helped make this work possible.
\end{acknowledgments}

\appendix


\begin{thebibliography}{38}%
\makeatletter
\providecommand \@ifxundefined [1]{%
 \@ifx{#1\undefined}
}%
\providecommand \@ifnum [1]{%
 \ifnum #1\expandafter \@firstoftwo
 \else \expandafter \@secondoftwo
 \fi
}%
\providecommand \@ifx [1]{%
 \ifx #1\expandafter \@firstoftwo
 \else \expandafter \@secondoftwo
 \fi
}%
\providecommand \natexlab [1]{#1}%
\providecommand \enquote  [1]{``#1''}%
\providecommand \bibnamefont  [1]{#1}%
\providecommand \bibfnamefont [1]{#1}%
\providecommand \citenamefont [1]{#1}%
\providecommand \href@noop [0]{\@secondoftwo}%
\providecommand \href [0]{\begingroup \@sanitize@url \@href}%
\providecommand \@href[1]{\@@startlink{#1}\@@href}%
\providecommand \@@href[1]{\endgroup#1\@@endlink}%
\providecommand \@sanitize@url [0]{\catcode `\\12\catcode `\$12\catcode
  `\&12\catcode `\#12\catcode `\^12\catcode `\_12\catcode `\%12\relax}%
\providecommand \@@startlink[1]{}%
\providecommand \@@endlink[0]{}%
\providecommand \url  [0]{\begingroup\@sanitize@url \@url }%
\providecommand \@url [1]{\endgroup\@href {#1}{\urlprefix }}%
\providecommand \urlprefix  [0]{URL }%
\providecommand \Eprint [0]{\href }%
\providecommand \doibase [0]{http://dx.doi.org/}%
\providecommand \selectlanguage [0]{\@gobble}%
\providecommand \bibinfo  [0]{\@secondoftwo}%
\providecommand \bibfield  [0]{\@secondoftwo}%
\providecommand \translation [1]{[#1]}%
\providecommand \BibitemOpen [0]{}%
\providecommand \bibitemStop [0]{}%
\providecommand \bibitemNoStop [0]{.\EOS\space}%
\providecommand \EOS [0]{\spacefactor3000\relax}%
\providecommand \BibitemShut  [1]{\csname bibitem#1\endcsname}%
\let\auto@bib@innerbib\@empty
\bibitem [{\citenamefont {Brandenburg}\ and\ \citenamefont
  {Lazarian}(2013)}]{Brandenburg2013}%
  \BibitemOpen
  \bibfield  {author} {\bibinfo {author} {\bibfnamefont {A.}~\bibnamefont
  {Brandenburg}}\ and\ \bibinfo {author} {\bibfnamefont {A.}~\bibnamefont
  {Lazarian}},\ }\href {\doibase 10.1007/s11214-013-0009-3} {\bibfield
  {journal} {\bibinfo  {journal} {Space Science Reviews}\ }\textbf {\bibinfo
  {volume} {178}},\ \bibinfo {pages} {163} (\bibinfo {year}
  {2013})}\BibitemShut {NoStop}%
\bibitem [{\citenamefont {Brandenburg}\ and\ \citenamefont
  {Subramanian}(2005)}]{Brandenburg2005}%
  \BibitemOpen
  \bibfield  {author} {\bibinfo {author} {\bibfnamefont {A.}~\bibnamefont
  {Brandenburg}}\ and\ \bibinfo {author} {\bibfnamefont {K.}~\bibnamefont
  {Subramanian}},\ }\href {\doibase 10.1016/j.physrep.2005.06.005} {\bibfield
  {journal} {\bibinfo  {journal} {Physics Reports}\ }\textbf {\bibinfo {volume}
  {417}},\ \bibinfo {pages} {1 } (\bibinfo {year} {2005})}\BibitemShut
  {NoStop}%
\bibitem [{\citenamefont {{Tobias}}, \citenamefont {{Cattaneo}},\ and\
  \citenamefont {{Boldyrev}}(2013)}]{Tobias2013}%
  \BibitemOpen
  \bibfield  {author} {\bibinfo {author} {\bibfnamefont {S.~M.}\ \bibnamefont
  {{Tobias}}}, \bibinfo {author} {\bibfnamefont {F.}~\bibnamefont
  {{Cattaneo}}}, \ and\ \bibinfo {author} {\bibfnamefont {S.}~\bibnamefont
  {{Boldyrev}}},\ }in\ \href {https://books.google.de/books?id=ge3Se\_lOUOoC}
  {\emph {\bibinfo {booktitle} {Ten Chapters in Turbulence}}},\ \bibinfo
  {editor} {edited by\ \bibinfo {editor} {\bibfnamefont {P.}~\bibnamefont
  {Davidson}}, \bibinfo {editor} {\bibfnamefont {Y.}~\bibnamefont {Kaneda}}, \
  and\ \bibinfo {editor} {\bibfnamefont {K.}~\bibnamefont {Sreenivasan}}}\
  (\bibinfo  {publisher} {Cambridge University Press},\ \bibinfo {year}
  {2013})\ Chap.~\bibinfo {chapter} {9}, pp.\ \bibinfo {pages}
  {351--404}\BibitemShut {NoStop}%
\bibitem [{\citenamefont {Brunetti}\ and\ \citenamefont
  {Jones}(2015)}]{Brunetti2015}%
  \BibitemOpen
  \bibfield  {author} {\bibinfo {author} {\bibfnamefont {G.}~\bibnamefont
  {Brunetti}}\ and\ \bibinfo {author} {\bibfnamefont {T.~W.}\ \bibnamefont
  {Jones}},\ }\enquote {\bibinfo {title} {Cosmic rays in galaxy clusters and
  their interaction with magnetic fields},}\ in\ \href {\doibase
  10.1007/978-3-662-44625-6_20} {\emph {\bibinfo {booktitle} {Magnetic Fields
  in Diffuse Media}}},\ \bibinfo {editor} {edited by\ \bibinfo {editor}
  {\bibfnamefont {A.}~\bibnamefont {Lazarian}}, \bibinfo {editor}
  {\bibfnamefont {M.~E.}\ \bibnamefont {de~Gouveia Dal~Pino}}, \ and\ \bibinfo
  {editor} {\bibfnamefont {C.}~\bibnamefont {Melioli}}}\ (\bibinfo  {publisher}
  {Springer Berlin Heidelberg},\ \bibinfo {address} {Berlin, Heidelberg},\
  \bibinfo {year} {2015})\ pp.\ \bibinfo {pages} {557--598}\BibitemShut
  {NoStop}%
\bibitem [{\citenamefont {Frisch}(1995)}]{Frisch1995}%
  \BibitemOpen
  \bibfield  {author} {\bibinfo {author} {\bibfnamefont {U.}~\bibnamefont
  {Frisch}},\ }\href@noop {} {\emph {\bibinfo {title} {Turbulence: The Legacy
  of AN Kolmogorov}}}\ (\bibinfo  {publisher} {Cambridge University Press},\
  \bibinfo {year} {1995})\BibitemShut {NoStop}%
\bibitem [{\citenamefont {Biskamp}(2003)}]{biskamp}%
  \BibitemOpen
  \bibfield  {author} {\bibinfo {author} {\bibfnamefont {D.}~\bibnamefont
  {Biskamp}},\ }\href@noop {} {\emph {\bibinfo {title} {Magnetohydrodynamic
  Turbulence, by Dieter Biskamp, pp.~310.~ISBN 0521810116.~Cambridge, UK:
  Cambridge University Press, September 2003.}}}\ (\bibinfo  {publisher}
  {Cambrdige University Press},\ \bibinfo {year} {2003})\BibitemShut {NoStop}%
\bibitem [{\citenamefont {Kraichnan}(1967)}]{Kraichnan1967}%
  \BibitemOpen
  \bibfield  {author} {\bibinfo {author} {\bibfnamefont {R.~H.}\ \bibnamefont
  {Kraichnan}},\ }\href {\doibase http://dx.doi.org/10.1063/1.1762301}
  {\bibfield  {journal} {\bibinfo  {journal} {Physics of Fluids}\ }\textbf
  {\bibinfo {volume} {10}},\ \bibinfo {pages} {1417} (\bibinfo {year}
  {1967})}\BibitemShut {NoStop}%
\bibitem [{\citenamefont {Alexakis}, \citenamefont {Mininni},\ and\
  \citenamefont {Pouquet}(2005)}]{Alexakis2005}%
  \BibitemOpen
  \bibfield  {author} {\bibinfo {author} {\bibfnamefont {A.}~\bibnamefont
  {Alexakis}}, \bibinfo {author} {\bibfnamefont {P.~D.}\ \bibnamefont
  {Mininni}}, \ and\ \bibinfo {author} {\bibfnamefont {A.}~\bibnamefont
  {Pouquet}},\ }\href {\doibase 10.1103/PhysRevE.72.046301} {\bibfield
  {journal} {\bibinfo  {journal} {Phys. Rev. E}\ }\textbf {\bibinfo {volume}
  {72}},\ \bibinfo {pages} {046301} (\bibinfo {year} {2005})}\BibitemShut
  {NoStop}%
\bibitem [{\citenamefont {Verma}(2004)}]{Verma2004}%
  \BibitemOpen
  \bibfield  {author} {\bibinfo {author} {\bibfnamefont {M.~K.}\ \bibnamefont
  {Verma}},\ }\href {\doibase http://dx.doi.org/10.1016/j.physrep.2004.07.007}
  {\bibfield  {journal} {\bibinfo  {journal} {Physics Reports}\ }\textbf
  {\bibinfo {volume} {401}},\ \bibinfo {pages} {229 } (\bibinfo {year}
  {2004})}\BibitemShut {NoStop}%
\bibitem [{\citenamefont {Debliquy}, \citenamefont {Verma},\ and\ \citenamefont
  {Carati}(2005)}]{Debliquy2005}%
  \BibitemOpen
  \bibfield  {author} {\bibinfo {author} {\bibfnamefont {O.}~\bibnamefont
  {Debliquy}}, \bibinfo {author} {\bibfnamefont {M.~K.}\ \bibnamefont {Verma}},
  \ and\ \bibinfo {author} {\bibfnamefont {D.}~\bibnamefont {Carati}},\ }\href
  {\doibase 10.1063/1.1867996} {\bibfield  {journal} {\bibinfo  {journal}
  {Physics of Plasmas}\ }\textbf {\bibinfo {volume} {12}},\ \bibinfo {eid}
  {042309} (\bibinfo {year} {2005}),\ 10.1063/1.1867996}\BibitemShut {NoStop}%
\bibitem [{\citenamefont {Mininni}(2011)}]{Mininni2011}%
  \BibitemOpen
  \bibfield  {author} {\bibinfo {author} {\bibfnamefont {P.~D.}\ \bibnamefont
  {Mininni}},\ }\href {\doibase 10.1146/annurev-fluid-122109-160748} {\bibfield
   {journal} {\bibinfo  {journal} {Annual Review of Fluid Mechanics}\ }\textbf
  {\bibinfo {volume} {43}},\ \bibinfo {pages} {377} (\bibinfo {year} {2011})},\
  \Eprint
  {http://arxiv.org/abs/http://dx.doi.org/10.1146/annurev-fluid-122109-160748}
  {http://dx.doi.org/10.1146/annurev-fluid-122109-160748} \BibitemShut
  {NoStop}%
\bibitem [{\citenamefont {Aluie}\ and\ \citenamefont
  {Eyink}(2010)}]{Aluie2010}%
  \BibitemOpen
  \bibfield  {author} {\bibinfo {author} {\bibfnamefont {H.}~\bibnamefont
  {Aluie}}\ and\ \bibinfo {author} {\bibfnamefont {G.~L.}\ \bibnamefont
  {Eyink}},\ }\href {\doibase 10.1103/PhysRevLett.104.081101} {\bibfield
  {journal} {\bibinfo  {journal} {Phys. Rev. Lett.}\ }\textbf {\bibinfo
  {volume} {104}},\ \bibinfo {pages} {081101} (\bibinfo {year}
  {2010})}\BibitemShut {NoStop}%
\bibitem [{\citenamefont {Teaca}, \citenamefont {Carati},\ and\ \citenamefont
  {Domaradzki}(2011)}]{Teaca2011}%
  \BibitemOpen
  \bibfield  {author} {\bibinfo {author} {\bibfnamefont {B.}~\bibnamefont
  {Teaca}}, \bibinfo {author} {\bibfnamefont {D.}~\bibnamefont {Carati}}, \
  and\ \bibinfo {author} {\bibfnamefont {J.~A.}\ \bibnamefont {Domaradzki}},\
  }\href {\doibase 10.1063/1.3661086} {\bibfield  {journal} {\bibinfo
  {journal} {Physics of Plasmas}\ }\textbf {\bibinfo {volume} {18}},\ \bibinfo
  {pages} {112307} (\bibinfo {year} {2011})},\ \Eprint
  {http://arxiv.org/abs/http://dx.doi.org/10.1063/1.3661086}
  {http://dx.doi.org/10.1063/1.3661086} \BibitemShut {NoStop}%
\bibitem [{\citenamefont {Graham}, \citenamefont {Cameron},\ and\ \citenamefont
  {Schüssler}(2010)}]{Graham2010}%
  \BibitemOpen
  \bibfield  {author} {\bibinfo {author} {\bibfnamefont {J.~P.}\ \bibnamefont
  {Graham}}, \bibinfo {author} {\bibfnamefont {R.}~\bibnamefont {Cameron}}, \
  and\ \bibinfo {author} {\bibfnamefont {M.}~\bibnamefont {Schüssler}},\
  }\href {http://stacks.iop.org/0004-637X/714/i=2/a=1606} {\bibfield  {journal}
  {\bibinfo  {journal} {The Astrophysical Journal}\ }\textbf {\bibinfo {volume}
  {714}},\ \bibinfo {pages} {1606} (\bibinfo {year} {2010})}\BibitemShut
  {NoStop}%
\bibitem [{\citenamefont {{Fromang, S.}}\ and\ \citenamefont {{Papaloizou,
  J.}}(2007)}]{FromangS.2007}%
  \BibitemOpen
  \bibfield  {author} {\bibinfo {author} {\bibnamefont {{Fromang, S.}}}\ and\
  \bibinfo {author} {\bibnamefont {{Papaloizou, J.}}},\ }\href {\doibase
  10.1051/0004-6361:20077942} {\bibfield  {journal} {\bibinfo  {journal}
  {A\&A}\ }\textbf {\bibinfo {volume} {476}},\ \bibinfo {pages} {1113}
  (\bibinfo {year} {2007})}\BibitemShut {NoStop}%
\bibitem [{\citenamefont {Salvesen}\ \emph {et~al.}(2014)\citenamefont
  {Salvesen}, \citenamefont {Beckwith}, \citenamefont {Simon}, \citenamefont
  {M.},\ and\ \citenamefont {Begelman}}]{Salvesen2014}%
  \BibitemOpen
  \bibfield  {author} {\bibinfo {author} {\bibfnamefont {G.}~\bibnamefont
  {Salvesen}}, \bibinfo {author} {\bibfnamefont {K.}~\bibnamefont {Beckwith}},
  \bibinfo {author} {\bibfnamefont {J.~B.}\ \bibnamefont {Simon}}, \bibinfo
  {author} {\bibfnamefont {O.~S.}\ \bibnamefont {M.}}, \ and\ \bibinfo {author}
  {\bibfnamefont {M.~C.}\ \bibnamefont {Begelman}},\ }\href {\doibase
  10.1093/mnras/stt2281} {\bibfield  {journal} {\bibinfo  {journal} {Monthly
  Notices of the Royal Astronomical Society}\ }\textbf {\bibinfo {volume}
  {438}},\ \bibinfo {pages} {1355} (\bibinfo {year} {2014})},\ \Eprint
  {http://arxiv.org/abs/http://mnras.oxfordjournals.org/content/438/2/1355.full.pdf+html}
  {http://mnras.oxfordjournals.org/content/438/2/1355.full.pdf+html}
  \BibitemShut {NoStop}%
\bibitem [{\citenamefont {Yang}\ \emph {et~al.}(2016)\citenamefont {Yang},
  \citenamefont {Shi}, \citenamefont {Wan}, \citenamefont {Matthaeus},\ and\
  \citenamefont {Chen}}]{Yang2016}%
  \BibitemOpen
  \bibfield  {author} {\bibinfo {author} {\bibfnamefont {Y.}~\bibnamefont
  {Yang}}, \bibinfo {author} {\bibfnamefont {Y.}~\bibnamefont {Shi}}, \bibinfo
  {author} {\bibfnamefont {M.}~\bibnamefont {Wan}}, \bibinfo {author}
  {\bibfnamefont {W.~H.}\ \bibnamefont {Matthaeus}}, \ and\ \bibinfo {author}
  {\bibfnamefont {S.}~\bibnamefont {Chen}},\ }\href {\doibase
  10.1103/PhysRevE.93.061102} {\bibfield  {journal} {\bibinfo  {journal} {Phys.
  Rev. E}\ }\textbf {\bibinfo {volume} {93}},\ \bibinfo {pages} {061102}
  (\bibinfo {year} {2016})}\BibitemShut {NoStop}%
\bibitem [{\citenamefont {Moll}\ \emph {et~al.}(2011)\citenamefont {Moll},
  \citenamefont {Graham}, \citenamefont {Pratt}, \citenamefont {Cameron},
  \citenamefont {Müller},\ and\ \citenamefont {Schüssler}}]{Moll_2011}%
  \BibitemOpen
  \bibfield  {author} {\bibinfo {author} {\bibfnamefont {R.}~\bibnamefont
  {Moll}}, \bibinfo {author} {\bibfnamefont {J.~P.}\ \bibnamefont {Graham}},
  \bibinfo {author} {\bibfnamefont {J.}~\bibnamefont {Pratt}}, \bibinfo
  {author} {\bibfnamefont {R.~H.}\ \bibnamefont {Cameron}}, \bibinfo {author}
  {\bibfnamefont {W.-C.}\ \bibnamefont {Müller}}, \ and\ \bibinfo {author}
  {\bibfnamefont {M.}~\bibnamefont {Schüssler}},\ }\href {\doibase
  10.1088/0004-637x/736/1/36} {\bibfield  {journal} {\bibinfo  {journal} {The
  Astrophysical Journal}\ }\textbf {\bibinfo {volume} {736}},\ \bibinfo {pages}
  {36} (\bibinfo {year} {2011})}\BibitemShut {NoStop}%
\bibitem [{\citenamefont {Kida}\ and\ \citenamefont {Orszag}(1990)}]{Kida1990}%
  \BibitemOpen
  \bibfield  {author} {\bibinfo {author} {\bibfnamefont {S.}~\bibnamefont
  {Kida}}\ and\ \bibinfo {author} {\bibfnamefont {S.~A.}\ \bibnamefont
  {Orszag}},\ }\href {\doibase 10.1007/BF01065580} {\bibfield  {journal}
  {\bibinfo  {journal} {Journal of Scientific Computing}\ }\textbf {\bibinfo
  {volume} {5}},\ \bibinfo {pages} {85} (\bibinfo {year} {1990})}\BibitemShut
  {NoStop}%
\bibitem [{\citenamefont {Simon}, \citenamefont {Hawley},\ and\ \citenamefont
  {Beckwith}(2009)}]{Simon2009}%
  \BibitemOpen
  \bibfield  {author} {\bibinfo {author} {\bibfnamefont {J.~B.}\ \bibnamefont
  {Simon}}, \bibinfo {author} {\bibfnamefont {J.~F.}\ \bibnamefont {Hawley}}, \
  and\ \bibinfo {author} {\bibfnamefont {K.}~\bibnamefont {Beckwith}},\ }\href
  {http://stacks.iop.org/0004-637X/690/i=1/a=974} {\bibfield  {journal}
  {\bibinfo  {journal} {The Astrophysical Journal}\ }\textbf {\bibinfo {volume}
  {690}},\ \bibinfo {pages} {974} (\bibinfo {year} {2009})}\BibitemShut
  {NoStop}%
\bibitem [{\citenamefont {Aluie}\ and\ \citenamefont
  {Eyink}(2009)}]{Aluie2009}%
  \BibitemOpen
  \bibfield  {author} {\bibinfo {author} {\bibfnamefont {H.}~\bibnamefont
  {Aluie}}\ and\ \bibinfo {author} {\bibfnamefont {G.~L.}\ \bibnamefont
  {Eyink}},\ }\href {\doibase 10.1063/1.3266948} {\bibfield  {journal}
  {\bibinfo  {journal} {Physics of Fluids}\ }\textbf {\bibinfo {volume} {21}},\
  \bibinfo {pages} {115108} (\bibinfo {year} {2009})},\ \Eprint
  {http://arxiv.org/abs/http://dx.doi.org/10.1063/1.3266948}
  {http://dx.doi.org/10.1063/1.3266948} \BibitemShut {NoStop}%
\bibitem [{\citenamefont {{Hunt}}, \citenamefont {{Wray}},\ and\ \citenamefont
  {{Moin}}(1988)}]{Hunt1988}%
  \BibitemOpen
  \bibfield  {author} {\bibinfo {author} {\bibfnamefont {J.~C.~R.}\
  \bibnamefont {{Hunt}}}, \bibinfo {author} {\bibfnamefont {A.~A.}\
  \bibnamefont {{Wray}}}, \ and\ \bibinfo {author} {\bibfnamefont
  {P.}~\bibnamefont {{Moin}}},\ }in\ \href@noop {} {\emph {\bibinfo {booktitle}
  {Studying Turbulence Using Numerical Simulation Databases, 2}}}\ (\bibinfo
  {year} {1988})\BibitemShut {NoStop}%
\bibitem [{\citenamefont {CARATI}\ \emph {et~al.}(2006)\citenamefont {CARATI},
  \citenamefont {DEBLIQUY}, \citenamefont {KNAEPEN}, \citenamefont {TEACA},\
  and\ \citenamefont {VERMA}}]{CARATI2006}%
  \BibitemOpen
  \bibfield  {author} {\bibinfo {author} {\bibfnamefont {D.}~\bibnamefont
  {CARATI}}, \bibinfo {author} {\bibfnamefont {O.}~\bibnamefont {DEBLIQUY}},
  \bibinfo {author} {\bibfnamefont {B.}~\bibnamefont {KNAEPEN}}, \bibinfo
  {author} {\bibfnamefont {B.}~\bibnamefont {TEACA}}, \ and\ \bibinfo {author}
  {\bibfnamefont {M.}~\bibnamefont {VERMA}},\ }\href {\doibase
  10.1080/14685240600774017} {\bibfield  {journal} {\bibinfo  {journal}
  {Journal of Turbulence}\ }\textbf {\bibinfo {volume} {7}},\ \bibinfo {pages}
  {N51} (\bibinfo {year} {2006})},\ \Eprint
  {http://arxiv.org/abs/10.1080/14685240600774017} {10.1080/14685240600774017}
  \BibitemShut {NoStop}%
\bibitem [{\citenamefont {Bryan}\ \emph {et~al.}(2014)\citenamefont {Bryan},
  \citenamefont {Norman}, \citenamefont {O'Shea}, \citenamefont {Abel},
  \citenamefont {Wise}, \citenamefont {Turk}, \citenamefont {Reynolds},
  \citenamefont {Collins}, \citenamefont {Wang}, \citenamefont {Skillman},
  \citenamefont {Smith}, \citenamefont {Harkness}, \citenamefont {Bordner},
  \citenamefont {hoon Kim}, \citenamefont {Kuhlen}, \citenamefont {Xu},
  \citenamefont {Goldbaum}, \citenamefont {Hummels}, \citenamefont {Kritsuk},
  \citenamefont {Tasker}, \citenamefont {Skory}, \citenamefont {Simpson},
  \citenamefont {Hahn}, \citenamefont {Oishi}, \citenamefont {So},
  \citenamefont {Zhao}, \citenamefont {Cen}, \citenamefont {Li},\ and\
  \citenamefont {Collaboration}}]{Enzo2013}%
  \BibitemOpen
  \bibfield  {author} {\bibinfo {author} {\bibfnamefont {G.~L.}\ \bibnamefont
  {Bryan}}, \bibinfo {author} {\bibfnamefont {M.~L.}\ \bibnamefont {Norman}},
  \bibinfo {author} {\bibfnamefont {B.~W.}\ \bibnamefont {O'Shea}}, \bibinfo
  {author} {\bibfnamefont {T.}~\bibnamefont {Abel}}, \bibinfo {author}
  {\bibfnamefont {J.~H.}\ \bibnamefont {Wise}}, \bibinfo {author}
  {\bibfnamefont {M.~J.}\ \bibnamefont {Turk}}, \bibinfo {author}
  {\bibfnamefont {D.~R.}\ \bibnamefont {Reynolds}}, \bibinfo {author}
  {\bibfnamefont {D.~C.}\ \bibnamefont {Collins}}, \bibinfo {author}
  {\bibfnamefont {P.}~\bibnamefont {Wang}}, \bibinfo {author} {\bibfnamefont
  {S.~W.}\ \bibnamefont {Skillman}}, \bibinfo {author} {\bibfnamefont
  {B.}~\bibnamefont {Smith}}, \bibinfo {author} {\bibfnamefont {R.~P.}\
  \bibnamefont {Harkness}}, \bibinfo {author} {\bibfnamefont {J.}~\bibnamefont
  {Bordner}}, \bibinfo {author} {\bibfnamefont {J.}~\bibnamefont {hoon Kim}},
  \bibinfo {author} {\bibfnamefont {M.}~\bibnamefont {Kuhlen}}, \bibinfo
  {author} {\bibfnamefont {H.}~\bibnamefont {Xu}}, \bibinfo {author}
  {\bibfnamefont {N.}~\bibnamefont {Goldbaum}}, \bibinfo {author}
  {\bibfnamefont {C.}~\bibnamefont {Hummels}}, \bibinfo {author} {\bibfnamefont
  {A.~G.}\ \bibnamefont {Kritsuk}}, \bibinfo {author} {\bibfnamefont
  {E.}~\bibnamefont {Tasker}}, \bibinfo {author} {\bibfnamefont
  {S.}~\bibnamefont {Skory}}, \bibinfo {author} {\bibfnamefont {C.~M.}\
  \bibnamefont {Simpson}}, \bibinfo {author} {\bibfnamefont {O.}~\bibnamefont
  {Hahn}}, \bibinfo {author} {\bibfnamefont {J.~S.}\ \bibnamefont {Oishi}},
  \bibinfo {author} {\bibfnamefont {G.~C.}\ \bibnamefont {So}}, \bibinfo
  {author} {\bibfnamefont {F.}~\bibnamefont {Zhao}}, \bibinfo {author}
  {\bibfnamefont {R.}~\bibnamefont {Cen}}, \bibinfo {author} {\bibfnamefont
  {Y.}~\bibnamefont {Li}}, \ and\ \bibinfo {author} {\bibfnamefont {T.~E.}\
  \bibnamefont {Collaboration}},\ }\href
  {http://stacks.iop.org/0067-0049/211/i=2/a=19} {\bibfield  {journal}
  {\bibinfo  {journal} {The Astrophysical Journal Supplement Series}\ }\textbf
  {\bibinfo {volume} {211}},\ \bibinfo {pages} {19} (\bibinfo {year}
  {2014})}\BibitemShut {NoStop}%
\bibitem [{\citenamefont {Stone}\ \emph {et~al.}(2008)\citenamefont {Stone},
  \citenamefont {Gardiner}, \citenamefont {Teuben}, \citenamefont {Hawley},\
  and\ \citenamefont {Simon}}]{Athena2008}%
  \BibitemOpen
  \bibfield  {author} {\bibinfo {author} {\bibfnamefont {J.~M.}\ \bibnamefont
  {Stone}}, \bibinfo {author} {\bibfnamefont {T.~A.}\ \bibnamefont {Gardiner}},
  \bibinfo {author} {\bibfnamefont {P.}~\bibnamefont {Teuben}}, \bibinfo
  {author} {\bibfnamefont {J.~F.}\ \bibnamefont {Hawley}}, \ and\ \bibinfo
  {author} {\bibfnamefont {J.~B.}\ \bibnamefont {Simon}},\ }\href
  {http://stacks.iop.org/0067-0049/178/i=1/a=137} {\bibfield  {journal}
  {\bibinfo  {journal} {The Astrophysical Journal Supplement Series}\ }\textbf
  {\bibinfo {volume} {178}},\ \bibinfo {pages} {137} (\bibinfo {year}
  {2008})}\BibitemShut {NoStop}%
\bibitem [{\citenamefont {Grete}\ \emph {et~al.}(2016)\citenamefont {Grete},
  \citenamefont {Vlaykov}, \citenamefont {Schmidt},\ and\ \citenamefont
  {Schleicher}}]{Grete2016a}%
  \BibitemOpen
  \bibfield  {author} {\bibinfo {author} {\bibfnamefont {P.}~\bibnamefont
  {Grete}}, \bibinfo {author} {\bibfnamefont {D.~G.}\ \bibnamefont {Vlaykov}},
  \bibinfo {author} {\bibfnamefont {W.}~\bibnamefont {Schmidt}}, \ and\
  \bibinfo {author} {\bibfnamefont {D.~R.~G.}\ \bibnamefont {Schleicher}},\
  }\href {\doibase 10.1063/1.4954304} {\bibfield  {journal} {\bibinfo
  {journal} {Physics of Plasmas}\ }\textbf {\bibinfo {volume} {23}},\ \bibinfo
  {eid} {062317} (\bibinfo {year} {2016}),\ 10.1063/1.4954304}\BibitemShut
  {NoStop}%
\bibitem [{\citenamefont {Wang}\ and\ \citenamefont {Abel}(2009)}]{Wang2009}%
  \BibitemOpen
  \bibfield  {author} {\bibinfo {author} {\bibfnamefont {P.}~\bibnamefont
  {Wang}}\ and\ \bibinfo {author} {\bibfnamefont {T.}~\bibnamefont {Abel}},\
  }\href {http://stacks.iop.org/0004-637X/696/i=1/a=96} {\bibfield  {journal}
  {\bibinfo  {journal} {The Astrophysical Journal}\ }\textbf {\bibinfo {volume}
  {696}},\ \bibinfo {pages} {96} (\bibinfo {year} {2009})}\BibitemShut
  {NoStop}%
\bibitem [{\citenamefont {Dedner}\ \emph {et~al.}(2002)\citenamefont {Dedner},
  \citenamefont {Kemm}, \citenamefont {Kröner}, \citenamefont {Munz},
  \citenamefont {Schnitzer},\ and\ \citenamefont {Wesenberg}}]{Dedner2002}%
  \BibitemOpen
  \bibfield  {author} {\bibinfo {author} {\bibfnamefont {A.}~\bibnamefont
  {Dedner}}, \bibinfo {author} {\bibfnamefont {F.}~\bibnamefont {Kemm}},
  \bibinfo {author} {\bibfnamefont {D.}~\bibnamefont {Kröner}}, \bibinfo
  {author} {\bibfnamefont {C.-D.}\ \bibnamefont {Munz}}, \bibinfo {author}
  {\bibfnamefont {T.}~\bibnamefont {Schnitzer}}, \ and\ \bibinfo {author}
  {\bibfnamefont {M.}~\bibnamefont {Wesenberg}},\ }\href {\doibase
  10.1006/jcph.2001.6961} {\bibfield  {journal} {\bibinfo  {journal} {Journal
  of Computational Physics}\ }\textbf {\bibinfo {volume} {175}},\ \bibinfo
  {pages} {645 } (\bibinfo {year} {2002})}\BibitemShut {NoStop}%
\bibitem [{\citenamefont {Stone}\ and\ \citenamefont
  {Gardiner}(2009)}]{Stone2009}%
  \BibitemOpen
  \bibfield  {author} {\bibinfo {author} {\bibfnamefont {J.~M.}\ \bibnamefont
  {Stone}}\ and\ \bibinfo {author} {\bibfnamefont {T.}~\bibnamefont
  {Gardiner}},\ }\href {\doibase https://doi.org/10.1016/j.newast.2008.06.003}
  {\bibfield  {journal} {\bibinfo  {journal} {New Astronomy}\ }\textbf
  {\bibinfo {volume} {14}},\ \bibinfo {pages} {139 } (\bibinfo {year}
  {2009})}\BibitemShut {NoStop}%
\bibitem [{\citenamefont {{Schmidt}}\ \emph {et~al.}(2009)\citenamefont
  {{Schmidt}}, \citenamefont {{Federrath}}, \citenamefont {{Hupp}},
  \citenamefont {{Kern}},\ and\ \citenamefont {{Niemeyer}}}]{Schmidt2009}%
  \BibitemOpen
  \bibfield  {author} {\bibinfo {author} {\bibfnamefont {W.}~\bibnamefont
  {{Schmidt}}}, \bibinfo {author} {\bibfnamefont {C.}~\bibnamefont
  {{Federrath}}}, \bibinfo {author} {\bibfnamefont {M.}~\bibnamefont {{Hupp}}},
  \bibinfo {author} {\bibfnamefont {S.}~\bibnamefont {{Kern}}}, \ and\ \bibinfo
  {author} {\bibfnamefont {J.~C.}\ \bibnamefont {{Niemeyer}}},\ }\href
  {\doibase 10.1051/0004-6361:200809967} {\bibfield  {journal} {\bibinfo
  {journal} {Astronomy \& Astrophysics}\ }\textbf {\bibinfo {volume} {494}},\
  \bibinfo {pages} {127} (\bibinfo {year} {2009})}\BibitemShut {NoStop}%
\bibitem [{\citenamefont {Kritsuk}\ \emph {et~al.}(2011)\citenamefont
  {Kritsuk}, \citenamefont {Åke Nordlund}, \citenamefont {Collins},
  \citenamefont {Padoan}, \citenamefont {Norman}, \citenamefont {Abel},
  \citenamefont {Banerjee}, \citenamefont {Federrath}, \citenamefont {Flock},
  \citenamefont {Lee}, \citenamefont {Li}, \citenamefont {Müller},
  \citenamefont {Teyssier}, \citenamefont {Ustyugov}, \citenamefont {Vogel},\
  and\ \citenamefont {Xu}}]{Kritsuk2011}%
  \BibitemOpen
  \bibfield  {author} {\bibinfo {author} {\bibfnamefont {A.~G.}\ \bibnamefont
  {Kritsuk}}, \bibinfo {author} {\bibnamefont {Åke Nordlund}}, \bibinfo
  {author} {\bibfnamefont {D.}~\bibnamefont {Collins}}, \bibinfo {author}
  {\bibfnamefont {P.}~\bibnamefont {Padoan}}, \bibinfo {author} {\bibfnamefont
  {M.~L.}\ \bibnamefont {Norman}}, \bibinfo {author} {\bibfnamefont
  {T.}~\bibnamefont {Abel}}, \bibinfo {author} {\bibfnamefont {R.}~\bibnamefont
  {Banerjee}}, \bibinfo {author} {\bibfnamefont {C.}~\bibnamefont {Federrath}},
  \bibinfo {author} {\bibfnamefont {M.}~\bibnamefont {Flock}}, \bibinfo
  {author} {\bibfnamefont {D.}~\bibnamefont {Lee}}, \bibinfo {author}
  {\bibfnamefont {P.~S.}\ \bibnamefont {Li}}, \bibinfo {author} {\bibfnamefont
  {W.-C.}\ \bibnamefont {Müller}}, \bibinfo {author} {\bibfnamefont
  {R.}~\bibnamefont {Teyssier}}, \bibinfo {author} {\bibfnamefont {S.~D.}\
  \bibnamefont {Ustyugov}}, \bibinfo {author} {\bibfnamefont {C.}~\bibnamefont
  {Vogel}}, \ and\ \bibinfo {author} {\bibfnamefont {H.}~\bibnamefont {Xu}},\
  }\href {http://stacks.iop.org/0004-637X/737/i=1/a=13} {\bibfield  {journal}
  {\bibinfo  {journal} {The Astrophysical Journal}\ }\textbf {\bibinfo {volume}
  {737}},\ \bibinfo {pages} {13} (\bibinfo {year} {2011})}\BibitemShut
  {NoStop}%
\bibitem [{\citenamefont {Porter}, \citenamefont {Jones},\ and\ \citenamefont
  {Ryu}(2015)}]{Porter2015}%
  \BibitemOpen
  \bibfield  {author} {\bibinfo {author} {\bibfnamefont {D.~H.}\ \bibnamefont
  {Porter}}, \bibinfo {author} {\bibfnamefont {T.~W.}\ \bibnamefont {Jones}}, \
  and\ \bibinfo {author} {\bibfnamefont {D.}~\bibnamefont {Ryu}},\ }\href
  {http://stacks.iop.org/0004-637X/810/i=2/a=93} {\bibfield  {journal}
  {\bibinfo  {journal} {The Astrophysical Journal}\ }\textbf {\bibinfo {volume}
  {810}},\ \bibinfo {pages} {93} (\bibinfo {year} {2015})}\BibitemShut
  {NoStop}%
\bibitem [{\citenamefont {Miesch}\ \emph {et~al.}(2015)\citenamefont {Miesch},
  \citenamefont {Matthaeus}, \citenamefont {Brandenburg}, \citenamefont
  {Petrosyan}, \citenamefont {Pouquet}, \citenamefont {Cambon}, \citenamefont
  {Jenko}, \citenamefont {Uzdensky}, \citenamefont {Stone}, \citenamefont
  {Tobias}, \citenamefont {Toomre},\ and\ \citenamefont {Velli}}]{Miesch2015}%
  \BibitemOpen
  \bibfield  {author} {\bibinfo {author} {\bibfnamefont {M.}~\bibnamefont
  {Miesch}}, \bibinfo {author} {\bibfnamefont {W.}~\bibnamefont {Matthaeus}},
  \bibinfo {author} {\bibfnamefont {A.}~\bibnamefont {Brandenburg}}, \bibinfo
  {author} {\bibfnamefont {A.}~\bibnamefont {Petrosyan}}, \bibinfo {author}
  {\bibfnamefont {A.}~\bibnamefont {Pouquet}}, \bibinfo {author} {\bibfnamefont
  {C.}~\bibnamefont {Cambon}}, \bibinfo {author} {\bibfnamefont
  {F.}~\bibnamefont {Jenko}}, \bibinfo {author} {\bibfnamefont
  {D.}~\bibnamefont {Uzdensky}}, \bibinfo {author} {\bibfnamefont
  {J.}~\bibnamefont {Stone}}, \bibinfo {author} {\bibfnamefont
  {S.}~\bibnamefont {Tobias}}, \bibinfo {author} {\bibfnamefont
  {J.}~\bibnamefont {Toomre}}, \ and\ \bibinfo {author} {\bibfnamefont
  {M.}~\bibnamefont {Velli}},\ }\href {\doibase 10.1007/s11214-015-0190-7}
  {\bibfield  {journal} {\bibinfo  {journal} {Space Science Reviews}\ }\textbf
  {\bibinfo {volume} {194}},\ \bibinfo {pages} {97} (\bibinfo {year}
  {2015})}\BibitemShut {NoStop}%
\bibitem [{\citenamefont {Vlaykov}\ \emph {et~al.}(2016)\citenamefont
  {Vlaykov}, \citenamefont {Grete}, \citenamefont {Schmidt},\ and\
  \citenamefont {Schleicher}}]{Vlaykov2016a}%
  \BibitemOpen
  \bibfield  {author} {\bibinfo {author} {\bibfnamefont {D.~G.}\ \bibnamefont
  {Vlaykov}}, \bibinfo {author} {\bibfnamefont {P.}~\bibnamefont {Grete}},
  \bibinfo {author} {\bibfnamefont {W.}~\bibnamefont {Schmidt}}, \ and\
  \bibinfo {author} {\bibfnamefont {D.~R.~G.}\ \bibnamefont {Schleicher}},\
  }\href {\doibase 10.1063/1.4954303} {\bibfield  {journal} {\bibinfo
  {journal} {Physics of Plasmas}\ }\textbf {\bibinfo {volume} {23}},\ \bibinfo
  {eid} {062316} (\bibinfo {year} {2016}),\ 10.1063/1.4954303}\BibitemShut
  {NoStop}%
\bibitem [{\citenamefont {Bardina}, \citenamefont {Ferziger},\ and\
  \citenamefont {Reynolds}(1980)}]{BARDINA1980}%
  \BibitemOpen
  \bibfield  {author} {\bibinfo {author} {\bibfnamefont {J.}~\bibnamefont
  {Bardina}}, \bibinfo {author} {\bibfnamefont {J.}~\bibnamefont {Ferziger}}, \
  and\ \bibinfo {author} {\bibfnamefont {W.}~\bibnamefont {Reynolds}}\
  }(\bibinfo  {publisher} {American Institute of Aeronautics and
  Astronautics},\ \bibinfo {year} {1980})\BibitemShut {NoStop}%
\bibitem [{\citenamefont {Grete}\ \emph {et~al.}(2017)\citenamefont {Grete},
  \citenamefont {Vlaykov}, \citenamefont {Schmidt},\ and\ \citenamefont
  {Schleicher}}]{Grete2017}%
  \BibitemOpen
  \bibfield  {author} {\bibinfo {author} {\bibfnamefont {P.}~\bibnamefont
  {Grete}}, \bibinfo {author} {\bibfnamefont {D.~G.}\ \bibnamefont {Vlaykov}},
  \bibinfo {author} {\bibfnamefont {W.}~\bibnamefont {Schmidt}}, \ and\
  \bibinfo {author} {\bibfnamefont {D.~R.~G.}\ \bibnamefont {Schleicher}},\
  }\href {\doibase 10.1103/PhysRevE.95.033206} {\bibfield  {journal} {\bibinfo
  {journal} {Phys. Rev. E}\ }\textbf {\bibinfo {volume} {95}},\ \bibinfo
  {pages} {033206} (\bibinfo {year} {2017})}\BibitemShut {NoStop}%
\bibitem [{\citenamefont {Aluie}(2013)}]{Aluie2013}%
  \BibitemOpen
  \bibfield  {author} {\bibinfo {author} {\bibfnamefont {H.}~\bibnamefont
  {Aluie}},\ }\href {\doibase http://dx.doi.org/10.1016/j.physd.2012.12.009}
  {\bibfield  {journal} {\bibinfo  {journal} {Physica D: Nonlinear Phenomena}\
  }\textbf {\bibinfo {volume} {247}},\ \bibinfo {pages} {54 } (\bibinfo {year}
  {2013})}\BibitemShut {NoStop}%
\bibitem [{\citenamefont {Domaradzki}, \citenamefont {Teaca},\ and\
  \citenamefont {Carati}(2010)}]{Domaradzki2010}%
  \BibitemOpen
  \bibfield  {author} {\bibinfo {author} {\bibfnamefont {J.~A.}\ \bibnamefont
  {Domaradzki}}, \bibinfo {author} {\bibfnamefont {B.}~\bibnamefont {Teaca}}, \
  and\ \bibinfo {author} {\bibfnamefont {D.}~\bibnamefont {Carati}},\ }\href
  {\doibase 10.1063/1.3431227} {\bibfield  {journal} {\bibinfo  {journal}
  {Physics of Fluids}\ }\textbf {\bibinfo {volume} {22}},\ \bibinfo {pages}
  {051702} (\bibinfo {year} {2010})},\ \Eprint
  {http://arxiv.org/abs/http://dx.doi.org/10.1063/1.3431227}
  {http://dx.doi.org/10.1063/1.3431227} \BibitemShut {NoStop}%
\end{thebibliography}
%

\end{document}